\documentclass[apj,twocolappendix, numberedappendixn]{emulateapj}
\usepackage[colorlinks,citecolor=blue]{hyperref}
\usepackage{textcomp}
\usepackage{subfigure}
\usepackage{verbatim}
\usepackage{bm}% bold math
\usepackage{hyperref}
\usepackage{amsmath}
\usepackage{graphicx}
\usepackage{epstopdf}
\usepackage{amssymb}
\usepackage{extarrows}
\usepackage{color}
\usepackage{CJK}
\usepackage{cancel}
\usepackage{ulem}
\usepackage[utf8]{inputenc}
\usepackage{url}
\usepackage{float}
\usepackage{mathtools}
\usepackage{lineno}

\newcommand{\ba}{\begin{eqnarray}}
\newcommand{\ea}{\end{eqnarray}}
\newcommand{\be}{\begin{equation}}
\newcommand{\ee}{\end{equation}}

\newcommand{\gw}{\mathrm{GW}}

\newcommand{\m}{\mathrm{max}}

\newcommand{\zlk}{\mathrm{ZLK}}

\newcommand{\s}{\mathrm{s}}

\newcommand{\bs}{\mathrm{bs}}
\newcommand{\Int}{\mathrm{int}}

\newcommand{\tide}{\mathrm{tide}}

\def\e1{e_1^2}

\def\2G{\scriptscriptstyle 2G}
\def\1G{\scriptscriptstyle 1G}
\def\3G{\scriptscriptstyle 3G}

\definecolor{ochre}{rgb}{0.8, 0.47, 0.13}

\begin{document}
\title{Hierarchical Black Hole Mergers in Nuclear Star Clusters: A Combined Dynamical-Secular Channel for GW231123-like Events}

%\author[0000-0002-0643-8295]{Bin Liu}
%\affiliation{Institute for Astronomy, School of Physics, Zhejiang University, 310027 Hangzhou, China}
%\affiliation{Center for Cosmology and Computational Astrophysics, Institute for Advanced Study in Physics, Zhejiang University, 310027 Hangzhou, China}
%\email{liubin23@zju.edu.cn}
%
%\author[0000-0002-1934-6250]{Dong Lai}
%\affiliation{Tsung-Dao Lee Institute, Shanghai Jiao Tong University, Shanghai 200240, China}
%\affiliation{Department of Astronomy, Center for Astrophysics and Planetary Science, Cornell University, Ithaca, NY 14853, USA}

\author{Bin Liu$^{1,2}$, Dieran Wang$^{3}$, Dong Lai$^{3,4}$}
\affil{$^{1}$ Institute for Astronomy, School of Physics, Zhejiang University, Hangzhou 310027 , China; liubin23@zju.edu.cn\\
$^{2}$ Center for Cosmology and Computational Astrophysics, Institute for Advanced Study in Physics, Zhejiang University, Hangzhou 310027, China\\
$^{3}$ Tsung-Dao Lee Institute, Shanghai Jiao Tong University, Shanghai 200240, China; donglai@sjtu.edu.cn\\
$^{4}$ Department of Astronomy, Center for Astrophysics and Planetary Science, Cornell University, Ithaca, NY 14853, USA
}

\begin{abstract}
The recent binary black hole (BH) merger GW231123, with both components likely in the high-mass gap and with high spins,
challenges standard BH binary formation models.
It is usually thought that the BHs are of second (or higher) generation (2G), resulting from the mergers of smaller BHs.
But the physical processes that produce the merging 2G BH binaries are unclear.
We suggest that such 2G mergers may be produced in the nuclear star cluster of a Milky Way-like galaxy.
A potentially dominant channel combines a sequence of binary-single interactions with secular evolution driven by the central supermassive BH.
Our model yields a merger rate that is consistent with the inferred rate of GW231123 within current uncertainties,
and further predicts an abundant population of 2G BH-star (or low-mass BH) binaries;
these binaries may observationally manifest as micro tidal disruption events or low-frequency gravitational-wave (GW) sources.
Detecting these binaries would provide crucial insights into the dynamical pathways of hierarchical BH assembly.
\end{abstract}

\section{Introduction}
\label{sec 1}

The fourth observing run of the LIGO-Virgo-KAGRA Collaboration has significantly expanded the GW transient catalog,
bringing the total number of detected compact binary mergers to about 400 \citep[e.g.,][]{LIGO-2025,LIGO-2026}.
While events like GW190521 provided initial evidence for hierarchical BH mergers \citep[e.g.,][]{GW190521},
the recent detection of GW231123 presents a more profound challenge to standard binary evolution models
\citep[e.g.,][]{GW231123,Lipunov-1997,Lipunov-2007,Podsiadlowski-2003,Belczynski-2010,Belczynski-2016,Dominik-2012,
Dominik-2013,Dominik-2015,Alejandro-2017}.
This BH binary (BHB) merger features components with masses $m_1=137^{+22}_{-17}M_\odot$ and $m_2=103^{+20}_{-52}M_\odot$
and spin parameters $\chi_1=0.9$ and $\chi_2=0.8$, indicating that both likely lie in the so-called ``high-mass gap".
The combination of high masses and high spins is inconsistent with first-generation (1G) BHs born from direct stellar collapse,
suggesting that GW231123 is a merger between two second-generation (2G) or higher generation BHs, each likely the remnant of a previous BH merger
\citep[e.g.,][]{Liu-HierarchicalMerger,Johan-2G,Fishbach-NA,Vigna-Gomez 2021,Li-PRL-2024,Stegmann-2G,Rasio-2G,
Arca Sedda-2G,McKernan-2G,Haiman-2G,Fishbach-2G,Metzger-2G}.

Nuclear star clusters (NSCs) and active galactic \mbox{nuclei} (AGN) disk, with their deep potential wells and high stellar densities,
are promising locations for such hierarchical growth
\citep[e.g.,][]{Baruteau-2011,McKernan-2012,Fishbach 2017,McKernan-2018,Bartos-2017,Stone-2017,Leigh-2018,Liu-2019,Secunda-2019,
Yang-2019,Liu-2020,Grobner-2020,Ishibashi-2020,Tagawa-2020,Tagawa-2021a,Tagawa-2021b,Tagawa-2021c,
Liyaping-2021,Ford-2021,Samsing-Nature,Lirixin-2022,Lijiaru-2022,Stegmann-2023,Yubo 2025,Mor AGN}.
These dense environments foster frequent dynamical encounters, and the high escape velocities aid in retaining 2G BHs.
While scattering process and secular effects are both critical in producing BHB mergers in NSCs,
previous studies have generally addressed them separately through parameter-heavy numerical simulations.
This makes it difficult to establish a clear picture of how bound 2G/2G BHBs form
or to identify the main pathways leading to BHB \mbox{mergers}
\citep[e.g.,][]{Zwart(2000),OLeary(2006),Miller(2009),Banerjee(2010),Downing(2010),Ziosi(2014),Rodriguez(2015),Rodriguez(2018),Samsing(2017),
Samsing(2018),Gondan(2018),Fragione 2023,Mor GC,Yubo 2025b}.

In this paper,
we explore whether the synergy between dynamical interactions and secular evolution may provide a pathway for
the formation and merger of 2G/2G BHBs in NSCs.
Using semi-analytical \mbox{models},
we find evidence for a two-step mechanism:
A sequence of binary-single interactions (stellar binary interacting with 2G BH,  and star/2G BH binary with another 2G BH)
first produce bound 2G/2G binaries.
These binaries can then merge rapidly either via GW emission or through the von Zeipel–Lidov–Kozai (ZLK) effect induced by the central SMBH
\citep[e.g.,][]{vonZeipel,Lidov,Kozai,Smadar,Miller-2002,Wen-2003,Antonini-2012,Antonini(2017),Silsbee(2017),Petrovich-2017,
Liu-ApJ,Liu-Quadruple,Xianyu-2018,Hoang-2018,Fragione-Quadruple,Fragione-nulearcluster,Zevin-2019}.
We emphasize that our model relies on several simplifying assumptions; a more detailed treatment is left for future work.

\section{Dynamical process}
\label{sec 2}

\begin{figure}
\centering
\begin{tabular}{cccc}
\includegraphics[width=8.5cm]{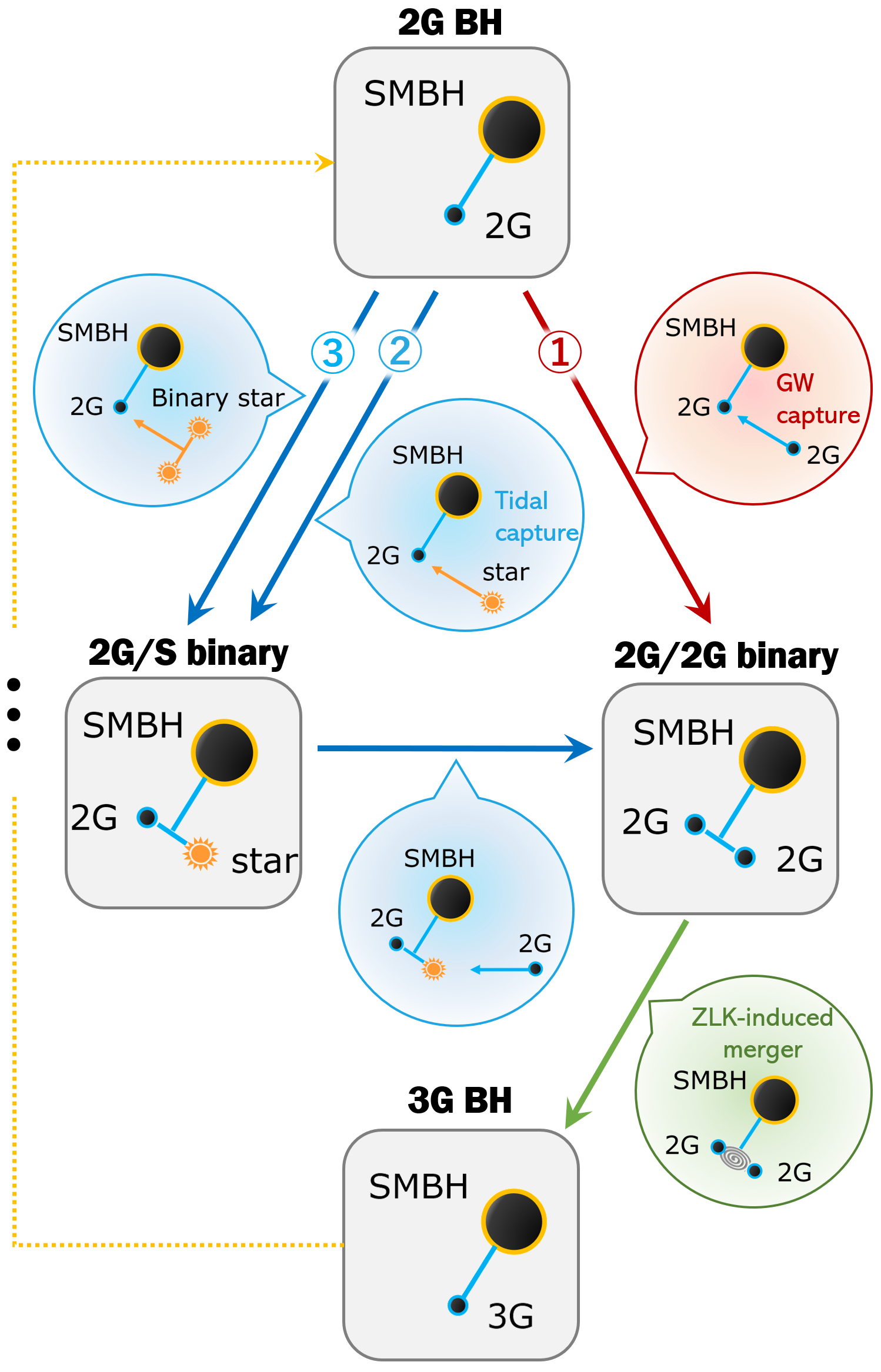}
\end{tabular}
\caption{Dynamical pathways for 2G/2G mergers in NSCs.
The evolution begins with a 2G BH (top), which may \mbox{form} a 2G/2G binary through GW capture (Channel 1).
Alternatively, it can first form a 2G/star (S) binary via tidal capture (Channel 2)
or binary–single (S/S binary-2G) interaction (Channel 3),
followed by an exchange interaction with another 2G BH that yields a 2G/2G binary.
This binary then merges into a 3G BH driven by secular forcing from the SMBH (ZLK effect),
potentially initiating further hierarchical growth.
}
\label{fig:schematic diagram}
\end{figure}

The evolution of BH systems in dense environments like NSCs involves complex dynamics,
governed by the SMBH's tidal gravity, multi-body scatterings and secular interactions \citep[e.g.,][]{Alexander 2017}.
To clarify the formation pathways of 2G/2G BHB mergers, we focus on three representative dynamical channels,
illustrated in Fig. \ref{fig:schematic diagram}, while leaving more complex interactions such as binary–binary scattering for future studies.

We model the NSC with four distinct populations.
The stellar component with mass $m_s$ ($\simeq1M_\odot$) follows a power-law density cusp \citep[e.g.,][]{BW-1976},
%%%%%%%%%%%%%%%%%%%%%%%%%%%%%%%%%%%%%%%%%%%%%%%%%%%%%%%%%%%%%%%%%%%%%%
\ba\label{eq:density profile}
n_s\equiv n_0\left(\frac{R}{\mathrm{pc}}\right)^{-\gamma}
= \frac{1.35\times10^6\ M_\odot}{\mathrm{pc}^3}\left(\frac{R}{0.25\ \mathrm{pc}}\right)^{-\gamma},
\ea
%%%%%%%%%%%%%%%%%%%%%%%%%%%%%%%%%%%%%%%%%%%%%%%%%%%%%%%%%%%%%%%%%%%%%%
where $n_0$ is the normalized density at distance $R=0.25$ pc and $\gamma$ is the slope of the density distribution.
This profile agrees with recent observational constraints for the Galactic center \citep[e.g.,][]{Genzel 2010}.
The differential number of stars within a shell at $R$ follows as
$dN_s = 4\pi n_0 R^{3-\gamma} \ d\ln R$.
The one-dimensional velocity dispersion, $\sigma_s = \sqrt{GM_\bullet/[R(1+\gamma)]}$,
is set by the SMBH of mass $M_\bullet$ \citep[e.g.,][]{Alexander 1999,Alexander 2014},
and the SMBH's influence radius $R_\mathrm{inf}=GM_\bullet/\sigma^2_s$.
Stellar binaries are included with a number density $n_b={\cal F}_b n_s$
(with fiducial value ${\cal F}_b=10\%$) and each binary has mass $\simeq2M_\odot$.
We assume steady populations of 1G and 2G BHs in the cluster.
The 1G BHs have masses below the high mass gap ($\lesssim 60M_\odot$)
and a fiducial number density $n_{\1G}={\cal F}_{\1G}n_s$, with ${\cal F}_{\1G}=1\%$.
The 2G BHs are assigned a characteristic mass $m_{\2G}\simeq100M_\odot$
and a fiducial density $n_{\2G}={\cal F}_{\2G}n_s$ where ${\cal F}_{\2G}=0.05\%$
\footnote{
The adopted values of ${\cal F}_{\1G}$ and ${\cal F}_{\2G}$ are estimated from the initial mass function and star formation history.
The actual values remain uncertain.
We have therefore tested how variations in these fractions affect the 2G/2G merger rate. The results are presented in Appendix \ref{App:Appendix B}.}
.
After mass segregation, all populations share similar velocity dispersions, $v_s\simeq v_b\simeq v_{\2G}\simeq \sigma_s$ at radius $R$.
This four-population model is idealized, serving as a proof-of-concept for the combined dynamical-secular channel.
It is applied to the core of a NSC, where mass segregation makes the dominance of these four specific components plausible
\citep[e.g.,][]{Sari 2022,Sari 2023,Sari 2025}.
While a continuous mass spectrum would be more realistic, and we ignore finer details such as distinct density profiles and other mass components,
this simplification enables a clear evaluation of the relative importance of key dynamical processes.
In this work, the notation ``$A/B$" denotes a binary with component $A$ and $B$, and ``$A–B$" indicates a dynamical encounter between $A$ and $B$.

\textit{Channel 1: Gravitational wave capture}.
The first channel is the dynamical formation of a 2G/2G binary through GW emission.
Consider two 2G BHs approach each other with velocity $v_{\2G}\simeq \sigma_s$ (at $R\simeq\infty$) and
the closest approach distance $r_\Int$. The radiated GW energy is
$\Delta E_\gw (r_\Int)\simeq(85\pi/12\sqrt{2})[G^{7/2}\mu_{\2G,\2G}^2(2m_{\2G})^{5/2}]/[c^5(r_\Int^\gw)^{7/2}]$.
To \mbox{form} a bound 2G/2G binary requires
$(1/2)\mu_{\2G,\2G} v_{\2G}^2-\Delta E_\gw (r_\Int)=E_b\equiv-Gm_{\2G}^2/(2a^\gw_{{\2G}/{\2G}})\lesssim0$.
To avoid disruption of the 2G/2G binary by subsequent stellar encounters requires $E_b\lesssim-m_s v_s^2/2$.
This implies that the 2G/2G binary has a characteristic semi-major axis
%%%%%%%%%%%%%%%%%%%%%%%%%%%%%%%%%%%%%%%%%%%%%%%%%%%%%%%%%%%%%%%%%%%%%%
\ba\label{eq:a2G GW}
a_{{\2G}/{\2G}}^\gw\simeq \frac{G m_{\2G}^2}{m_s v_s^2}.
\ea
%%%%%%%%%%%%%%%%%%%%%%%%%%%%%%%%%%%%%%%%%%%%%%%%%%%%%%%%%%%%%%%%%%%%%%
The required interaction distance $r_\Int^\gw$ is set by $\Delta E_\gw (r_\Int)\simeq m_s v_s^2$,
giving $r_\Int^\gw= (85\pi/12)^{2/7} [m_{\2G}c^2/(m_s v_s^2)]^{2/7}(Gm_{\2G}/c^2)$.
The corresponding cross section for binary formation is
$\sigma_{{\2G}-{\2G}}\simeq \pi (v_\Int r_\Int^\gw/v_s)^2$, where $v_\Int\simeq (4G m_{\2G}/r_\Int^\gw)^{1/2}$.
The timescale for 2G/2G binary formation is
%%%%%%%%%%%%%%%%%%%%%%%%%%%%%%%%%%%%%%%%%%%%%%%%%%%%%%%%%%%%%%%%%%%%%%
\ba
T_{{\2G}-{\2G}}\simeq v_s(4\pi G m_{\2G}n_{\2G} r_\Int^\gw)^{-1}.
\ea
%%%%%%%%%%%%%%%%%%%%%%%%%%%%%%%%%%%%%%%%%%%%%%%%%%%%%%%%%%%%%%%%%%%%%%
The captured 2G/2G binary has an eccentricity given by $a_{{\2G}/{\2G}}^\gw(1-e_{{\2G}/{\2G}}^\gw)\simeq r_\Int^\gw$,
and merges quickly \citep[e.g.,][]{Peters}.

\textit{Channel 2: Tidal capture}.
A 2G BH may form a binary with another 2G BH through a two-step process.
It first tidally captures a star to form an intermediate 2G/star (2G/S) binary.
The timescale for this 2G–S encounter is
%%%%%%%%%%%%%%%%%%%%%%%%%%%%%%%%%%%%%%%%%%%%%%%%%%%%%%%%%%%%%%%%%%%%%%
\ba
T_{\2G-S}\simeq v_s( 2\pi G m_{\2G}n_\s r_\Int^\tide)^{-1},
\ea
%%%%%%%%%%%%%%%%%%%%%%%%%%%%%%%%%%%%%%%%%%%%%%%%%%%%%%%%%%%%%%%%%%%%%%
where $r_\Int^\tide$ is the pericenter distance for \mbox{significant} tidal energy dissipation ($\Delta E_\tide$).
We adopt $\Delta E_\tide\simeq(Gm_{\2G}^2/R_s)(R_s/r_\Int^\tide)^6(A\varepsilon^{-\beta})$
with $A=0.24$, $\beta=3.1$ appropriate for a polytrope of index $4/3$, $R_s$ the stellar radius and
$\varepsilon\simeq(m_s/m_{\2G})^{1/2}(r_\Int^\tide/R_s)^{3/2}$
\citep[e.g.,][]{Portegies TC,Lai TC}.
Setting $\Delta E_\tide\simeq m_s v_s^2$,
we have $r_\Int^\tide\simeq[AG m_s/(v_s^2 R_s)]^xR_s(m_{\2G}/m_s)^y$ with $x=2/(12+3\beta)$ and $y=(4+\beta)/(12+3\beta)$.

A bound, hard 2G/S binary forms if its binding energy satisfies
$|Gm_{\2G}m_s/(2a^\tide_{\2G/S})|\gtrsim m_s v_s^2/2$,
yielding a characteristic semi-major axis
%%%%%%%%%%%%%%%%%%%%%%%%%%%%%%%%%%%%%%%%%%%%%%%%%%%%%%%%%%%%%%%%%%%%%%
\ba
a_{\2G/S}^\tide\simeq \frac{Gm_{\2G}}{v_s^2}.
\ea
%%%%%%%%%%%%%%%%%%%%%%%%%%%%%%%%%%%%%%%%%%%%%%%%%%%%%%%%%%%%%%%%%%%%%%
Gravitational scatterings with other stars then harden the binary to
$a_{\2G/S}^{' \tide}\simeq \zeta a_{\2G/S}^\tide$,
where $\zeta$ is the hardening factor (we adopt the fiducial value $\zeta=0.5$ unless otherwise stated).
This 2G/S binary may further experience a binary–single exchange interaction
when another 2G BH approaches within a distance $\sim a_{\2G/S}^{' \tide}$.
The encounter timescale is:
%%%%%%%%%%%%%%%%%%%%%%%%%%%%%%%%%%%%%%%%%%%%%%%%%%%%%%%%%%%%%%%%%%%%%%
\ba\label{eq:T 2G/S-2G}
T_{\2G/S-\2G}\simeq v_s( 4\pi G m_{\2G}n_{\2G}a_{\2G/S}^{' \tide})^{-1}.
\ea
%%%%%%%%%%%%%%%%%%%%%%%%%%%%%%%%%%%%%%%%%%%%%%%%%%%%%%%%%%%%%%%%%%%%%%
During the interaction, an energy exchange of order $\Delta E\simeq Gm_{\2G}m_s/a_{\2G/S}^{' \tide}$ occurs.
If the lighter star is \mbox{ejected} with sufficiently large energy, a 2G/2G binary forms.
The binding energy of the new binary satisfies $Gm_{\2G}^2/(2a^\tide_{\2G/\2G})\simeq\Delta E/2$,
giving the characteristic semi-major axis
%%%%%%%%%%%%%%%%%%%%%%%%%%%%%%%%%%%%%%%%%%%%%%%%%%%%%%%%%%%%%%%%%%%%%%
\ba
a_{\2G/\2G}^\tide\simeq\eta a_{\2G/S}^{' \tide}\bigg(\frac{m_{\2G}}{m_s}\bigg)=\zeta\eta\frac{Gm_{\2G}^2}{m_s v_s^2},
\ea
%%%%%%%%%%%%%%%%%%%%%%%%%%%%%%%%%%%%%%%%%%%%%%%%%%%%%%%%%%%%%%%%%%%%%%
where the coefficient $\eta$ can be calibrated using N-body numerical integrations.
The corresponding eccentricity follows from $a_{\2G/\2G}^\tide(1-e_{\2G/\2G}^\tide)\simeq a_{\2G/S}^{' \tide}$
(see Appendix~\ref{App:Appendix A} for more details).
The resulting 2G/2G binary is relatively soft that could be disrupted by subsequent scatterings,
with an evaporation timescale
\footnote{
In our model, $T_\mathrm{evap}\simeq (m_{\2G}/m_s)^2 T_{\2G/\2G-S}$, in which $T_{\2G/\2G-S}$
denotes the timescale for a 2G/2G binary to interact with a field star, i.e.,
$T_{\2G/\2G-S}=v_s(4\pi G m_{\2G}n_s a_{\2G/\2G})^{-1}$.
}
\citep[e.g.,][]{Galactic dynamics}:
%%%%%%%%%%%%%%%%%%%%%%%%%%%%%%%%%%%%%%%%%%%%%%%%%%%%%%%%%%%%%%%%%%%%%%
\ba\label{eq:T evap}
T_\mathrm{evap}\simeq v_s
(4\pi G m_{\2G}n_s a_{\2G/\2G}^\tide)^{-1}(m_{\2G}/m_s)^2.
\ea
%%%%%%%%%%%%%%%%%%%%%%%%%%%%%%%%%%%%%%%%%%%%%%%%%%%%%%%%%%%%%%%%%%%%%%
Only the binaries that can merge within $T_\mathrm{evap}$ are taken into account in our rate estimate (see below).

\textit{Channel 3: Binary-single interaction}.
A third channel for 2G/2G mergers involves binary–single (BS) interaction between a 2G BH and a stellar binary.
Consider stellar binaries with semi-major axis $a_{s/s}\simeq Gm_s/v_s^2$.
The binary-2G encounter timescale is
%%%%%%%%%%%%%%%%%%%%%%%%%%%%%%%%%%%%%%%%%%%%%%%%%%%%%%%%%%%%%%%%%%%%%%
\ba
T_{\2G-S/S}\simeq v_s(2\pi G m_{\2G}n_\bs r_\Int^\mathrm{BS})^{-1},
\ea
%%%%%%%%%%%%%%%%%%%%%%%%%%%%%%%%%%%%%%%%%%%%%%%%%%%%%%%%%%%%%%%%%%%%%%
with the interaction distance $r_\Int^\mathrm{BS}\simeq a_{s/s}[m_{\2G}/(2m_s)]^{1/3}$.

This tidal disruption process, analogous to the Hill mechanism, releases energy
$\Delta E\simeq Gm_{\2G}m_s a_{s/s}/\big(r_\Int^\mathrm{BS}\big)^2$,
typically producing a bound 2G/S binary and ejecting the other star.
The 2G/S binary's semi-major axis can be evaluated by $Gm_{\2G}m_s/(2a^\mathrm{BS}_{\2G/S})\simeq\Delta E/2$
(\citealp{Yu fangyuan}; see also \citealp{Ginat 2021}):
%%%%%%%%%%%%%%%%%%%%%%%%%%%%%%%%%%%%%%%%%%%%%%%%%%%%%%%%%%%%%%%%%%%%%%
\ba
a_{\2G/S}^\mathrm{BS}\simeq a_{s/s}\bigg(\frac{m_{\2G}}{2m_s}\bigg)^{2/3}
=\frac{G m_s}{v_s^2}\bigg(\frac{m_{\2G}}{2m_s}\bigg)^{2/3}.
\ea
%%%%%%%%%%%%%%%%%%%%%%%%%%%%%%%%%%%%%%%%%%%%%%%%%%%%%%%%%%%%%%%%%%%%%%
Again, stellar scattering hardens the binary by a factor $\zeta$:
$a_{\2G/S}^{' \mathrm{BS}}\simeq \zeta a_{\2G/S}^\mathrm{BS}$.
A subsequent encounter with another 2G BH, occurring over a timescale
$T_{\2G/S-\2G}$ (Eq. \ref{eq:T 2G/S-2G} with $a_{\2G/S}^{' \tide}\rightarrow a_{\2G/S}^{' \mathrm{BS}}$),
has a finite probability of forming a 2G/2G binary. The resulting binary has a characteristic semi-major axis:
%%%%%%%%%%%%%%%%%%%%%%%%%%%%%%%%%%%%%%%%%%%%%%%%%%%%%%%%%%%%%%%%%%%%%%
\ba\label{eq:a 2G2G BS}
a_{\2G/\2G}^\mathrm{BS}\simeq\eta a_{\2G/S}^{' \mathrm{BS}}\bigg(\frac{m_{\2G}}{m_s}\bigg)
=\zeta\eta\bigg[\frac{Gm_{\2G}}{v_s^2}\bigg(\frac{m_{\2G}}{2m_s}\bigg)^{2/3}\bigg].
\ea
%%%%%%%%%%%%%%%%%%%%%%%%%%%%%%%%%%%%%%%%%%%%%%%%%%%%%%%%%%%%%%%%%%%%%%
Here, we set $\eta\simeq1/3$ based on the peak of the post-encounter semi-major axis distribution from our N-body simulations
(see Appendix \ref{App:Appendix A}).
\footnote{
Binary-single encounters can produce multiple outcomes including exchange, flyby, ionization/disruption, and physical collisions.
We have performed a series of N-body scattering simulations to calibrate the formation channel.
For 2G/S-2G encounters, we find a high probability of forming a 2G/2G binary.
The resulting distribution of the post-encounter orbital parameters is presented in Appendix \ref{App:Appendix A},
where we also show that the peak of the semi-major axis distribution yields $\eta \simeq 1/3$ across different hardening factors.
}
As before, the binary may be disrupted by later encounters; only those that merge
within $T_\mathrm{evap}$ (Eq. \ref{eq:T evap} with $a_{\2G/\2G}^{\tide}\rightarrow a_{\2G/\2G}^\mathrm{BS}$) contribute to the detection rate.

\begin{figure}
\centering
\begin{tabular}{cccc}
\includegraphics[width=8cm]{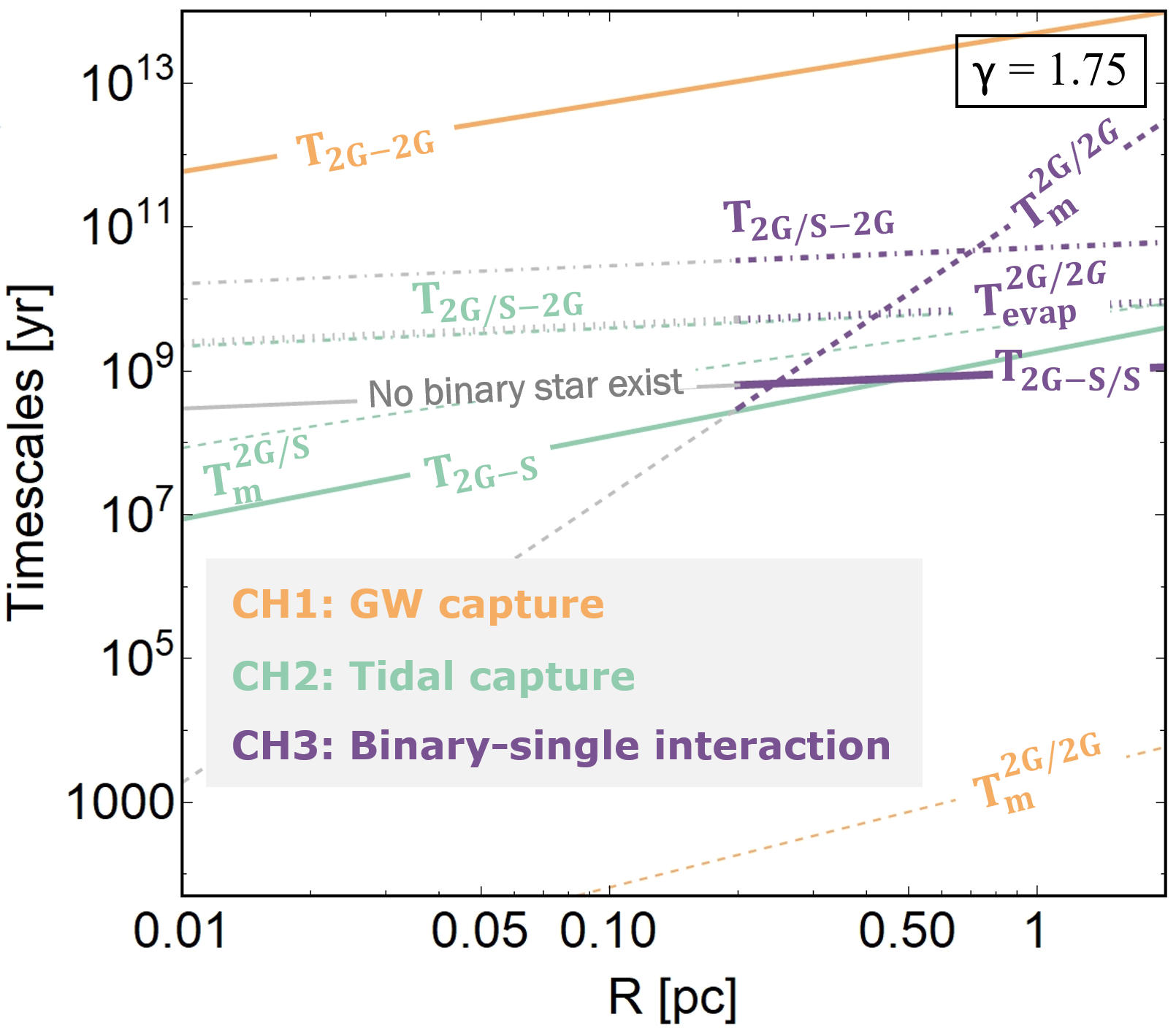}
\end{tabular}
\caption{Key timescales for forming 2G/2G mergers via different channels, evaluated at distance $R$ from an SMBH of mass
$M_\bullet=4\times10^6M_\odot$, assuming a density profile index $\gamma$ constant for all populations.
The merger time of various binaries due to GW emission (without the influence of the SMBH) are denoted by $T_\mathrm{m}$.}
\label{fig:Timescale}
\end{figure}

Fig. \ref{fig:Timescale} compares the timescales for 2G/2G binary formation through different channels.
In the GW capture channel (orange), the 2G–2G encounter timescale is long, but once formed, the binary merges rapidly.
For tidal capture (green), the ratio $T_\mathrm{m}^{\2G/S}/T_{\2G/S-\2G}\ll1$
indicates that 2G/S binaries merge before encountering another 2G BH.
These mergers may potentially produce micro tidal disruption events, making this channel ineffective for 2G/2G binary production.
The binary–single channel operates mainly at larger galactic distance $R$, where stellar binaries survive
($a_{s/s}\gtrsim2R_s$).
A newly formed 2G/S binary may undergo an exchange interaction to form a 2G/2G binary,
which subsequently merges efficiently ($T_\mathrm{m}/T_\mathrm{evap}\lesssim1$), thereby avoiding disruption by subsequent scatterings.

Note that including the ZLK effect from the SMBH can increase the eccentricity and accelerate mergers,
with $T_\mathrm{m}^\zlk\simeq T_{\mathrm{m},0}(1-e_\m^2)^3$, where
$T_{\mathrm{m},0}\equiv[5c^5 a_{\2G/\2G}^4]/[256 G^3 (2m_{\2G/\2G})^2 \mu_{\2G,\2G}]$
and $e_\m$ is the maximum eccentricity attained through ZLK oscillations
\footnote{
To account for the ZLK effect, we model the 2G/2G binary as moving on a moderately eccentric orbit (eccentricity $\sim$0.5) around the central SMBH.
}
\citep[e.g.,][]{Liu et al 2015,Liu-ApJ,Liu-HierarchicalMerger}.
Our analysis includes both GW-driven orbital decay and ZLK-induced merger mechanisms across all channels.
For the binaries shown in Fig. \ref{fig:Timescale}, the small semi-major axis strongly suppresses the ZLK effect.
However, for 2G/2G BHBs formed through binary–single scattering
(which have a broad semi-major axis distribution),
the ZLK effect may become efficient if we adopt values from the upper tail of the semi-major axis distribution
rather than its peak (see Appendix \ref{App:Appendix B}).

\section{Merger rate}
\label{sec 3}

To quantitatively assess each channel's contribution to 2G/2G mergers,
we formulate a multi-stage evolutionary framework,
where 2G BHs evolve sequentially through pairing and merger phases.
We note that the following rate calculation is primarily designed to elucidate the relative efficiencies of different channels
rather than to deliver absolute empirical predictions.
For the tidal capture and binary–single interaction channels,
we track four ``intermediate populations": single 2G BHs ($N_{\2G}$), 2G/S binaries ($N_{\2G/S}$), 2G/2G binaries ($N_{\2G/\2G}$),
and third-generation (3G) BHs ($N_{\3G}$).
For a typical galaxy, their evolution is governed by a set of equations:
%%%%%%%%%%%%%%%%%%%%%%%%%%%%%%%%%%%%%%%%%%%%%%%%%%%%%%%%%%%%%%%%%%%%%%
\begin{equation}\label{eq:Bateman equations}
\begin{cases}
\frac{dN_{2G/S}}{dt}=f_1\frac{N_{2G}}{T_A}-\frac{N_{2G/S}}{T_B}\\
\frac{dN_{2G/2G}}{dt}=f_2\frac{N_{2G/S}}{T_B}-\frac{N_{2G/2G}}{T_{\mathrm{m}}}\\
\frac{dN_{3G}}{dt}=f_3\frac{N_{2G/2G}}{T_{\mathrm{m}}}.
\end{cases}
\end{equation}
%%%%%%%%%%%%%%%%%%%%%%%%%%%%%%%%%%%%%%%%%%%%%%%%%%%%%%%%%%%%%%%%%%%%%%
Here, $T_A\equiv T_{\2G-S}$(tidal capture) or $T_{\2G-S/S}$ (binary single interaction), $T_B\equiv T_{\2G/S-\2G}$;
$f_1$ represents the probability of forming a 2G/S binary following a 2G–S/S interaction,
$f_2$ quantifies the probability of forming a 2G/2G binary via a subsequent 2G/S–2G encounter, and
$f_3$ accounts for the fraction of 2G/2G binaries that successfully merge within $T_\mathrm{evap}$.
We assume that there is a steady production of 2G BHs, so that $N_{\2G}=$ const.
In steady state, the merger rate $\Gamma=dN_{\3G}/dt$ at a given distance $R$ can be obtained analytically.
See Appendix \ref{App:Appendix B} for details.

\begin{figure}
\centering
\begin{tabular}{cccc}
\includegraphics[width=8cm]{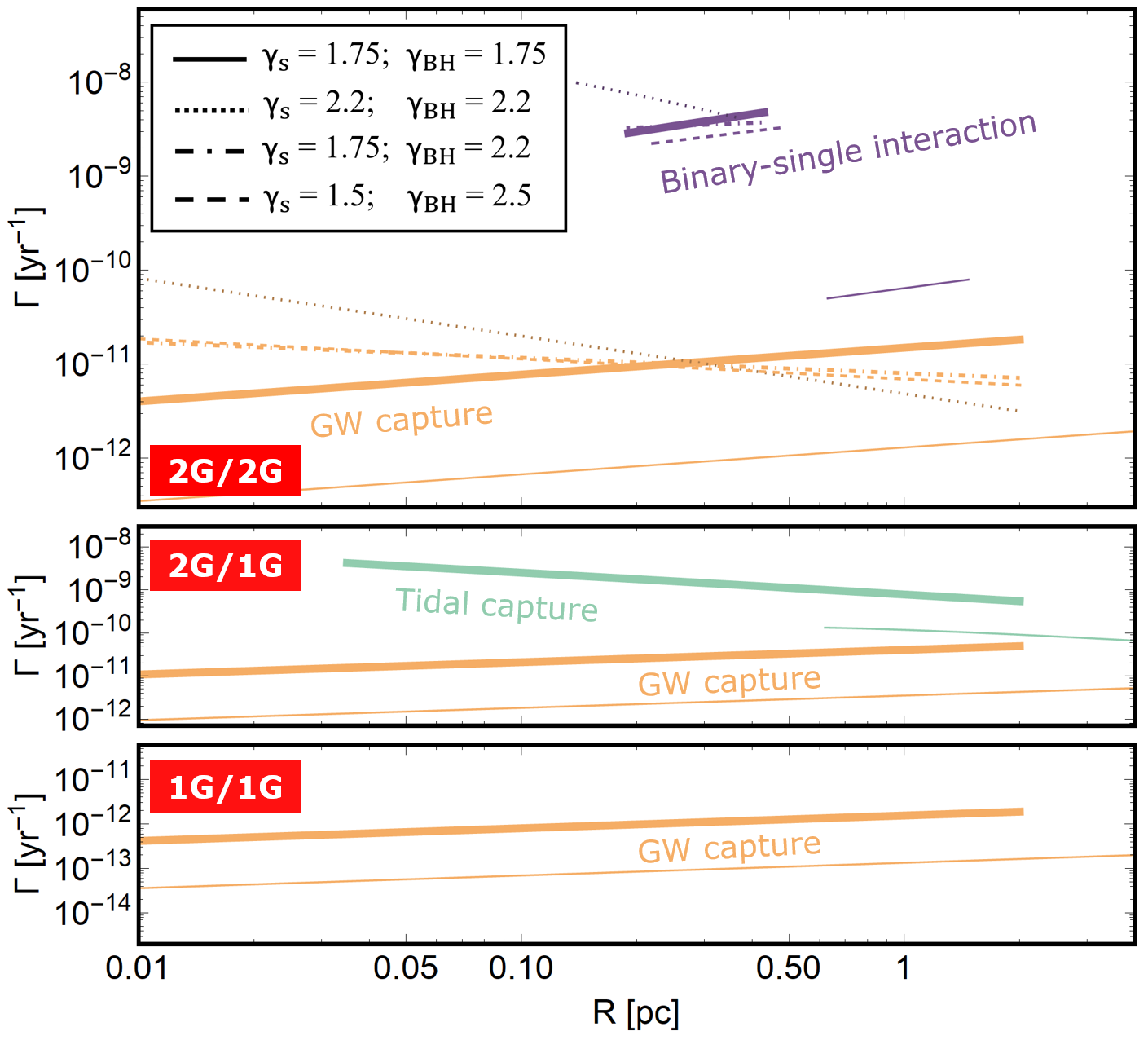}
\end{tabular}
\caption{Top panel: 2G/2G BHB merger rates versus the galactic distance $R$.
The rates are derived per galaxy from Eq.~(\ref{eq:Bateman equations}).
Different slopes of the density profile for star ($\gamma_s$) and BH ($\gamma_\mathrm{BH}$) are taken into account
(as labeled; see Eq. \ref{eq:density profile}).
We adopt $T_\mathrm{m}= T_\mathrm{evap}$ in Eq.~(\ref{eq:Bateman equations}).
The formation probability $f_1$ is $40\%$ for 2G/S binaries resulting from 2G–S/S encounters based on our N-body simulations,
and is set to $100\%$ for 2G/S binaries formed via tidal capture.
The formation probability $f_2$ is $30\%$ for 2G/2G BHBs based on the N-body simulations
and is assumed to be $10\%$ for 2G/1G and 1G/1G BHBs.
The parameter $f_3$ denotes the fraction of 2G/2G binaries that merge within $T_\mathrm{evap}$:
specifically, $f_3 = 100\%$ if $T_\mathrm{m}< T_\mathrm{m}^\zlk$ (signifying mergers via GW emission alone);
otherwise, $f_3$ represents the ZLK-driven fraction of systems that attain sufficiently high eccentricities to merge within $T_\mathrm{evap}$,
assuming isotropic orbital orientations \citep[e.g.,][]{Liu-HierarchicalMerger}.
Middle/Bottom panels: 2G/1G and 1G/1G merger rates for $\gamma=1.75$. Thick solid lines correspond to $M_\bullet=4\times10^6M_\odot$
and thin solid lines represent $M_\bullet=10^7M_\odot$.
}
\label{fig:Merger rate}
\end{figure}

The top panel of Fig. \ref{fig:Merger rate} shows the merger rates for the GW capture and binary–single channels.
Tidal capture is excluded since 2G/S binaries merge before encountering another 2G BH (Fig. \ref{fig:Timescale}).
These rates depend sensitively on $\gamma$,
which sets the stellar density $n_s$ and encounter timescales.
The GW capture channel is inefficient, due to the long 2G–2G encounter timescale $T_{{\2G}-{\2G}}$,
with the rate $10^2\sim 10^3$ times lower than the binary–single channel.
For the binary–single interaction, we exclude the distance ranges where stellar binaries
cannot survive or where the 2G/S binary merges within $T_{\2G/S-\2G}$.
For the remaining 2G/2G binaries, their high eccentricities enable rapid merger via GW emission alone.
However, systems with $R\gtrsim0.5$ pc have $T_\mathrm{m}>T_\mathrm{evap}$ and thus do not contribute to the merger rate
(see also Fig. \ref{fig:Timescale}).

Since the 2G/2G BHBs formed via binary–single scattering exhibit a broad distribution of semi-major axes $a_{\2G/\2G}^\mathrm{BS}$,
(see Appendix \ref{App:Appendix A}), we also consider the extreme values of the distribution
to examine the impact on the merger rate.
For the minimum value of $a_{\2G/\2G}^\mathrm{BS}$, the resulting merger behavior mirrors that of the peak case (shown in Fig. \ref{fig:Merger rate}).
For the maximum value, the ZLK mechanism plays a pivotal role in accelerating mergers,
with a fraction $f_3\lesssim15\%$ satisfying $T_\mathrm{m}^\zlk\lesssim T_\mathrm{evap}$.
Overall, as an illustrative estimate,
when scaled to the number of Milky Way-like galaxies in the local universe \citep[e.g.,][]{SDSS 2003,Eagle 2014},
the binary–single encounters can yield a merger rate that is consistent
with the inferred rate of GW231123 within the uncertainties of our model (See more details in Appendix \ref{App:Appendix B}).

We now consider the 1G BH population in NSCs.
To distinguish it from the 2G components, we assign a characteristic mass of $m_{\1G}=10M_\odot$ in the following calculations.
For the GW capture channel, the 2G/1G merger rate is computed as in the 2G/2G case (See Appendix \ref{App:Appendix C}),
giving a characteristic semi-major axis for 2G/1G binaries:
%%%%%%%%%%%%%%%%%%%%%%%%%%%%%%%%%%%%%%%%%%%%%%%%%%%%%%%%%%%%%%%%%%%%%%
\ba\label{eq:a 2G1G GW}
a_{{\2G}/{\1G}}^\gw\simeq\frac{Gm_{\2G}m_{\1G}}{m_s v_s^2}.
\ea
%%%%%%%%%%%%%%%%%%%%%%%%%%%%%%%%%%%%%%%%%%%%%%%%%%%%%%%%%%%%%%%%%%%%%%
For the tidal capture and binary–single channels,
we apply a similar analysis but replace the interaction with a second 2G BH by that with a 1G BH.
A 2G/S binary may thus undergo an exchange interaction with a 1G BH,
forming a 2G/1G binary. The characteristic semi-major axes are:
%%%%%%%%%%%%%%%%%%%%%%%%%%%%%%%%%%%%%%%%%%%%%%%%%%%%%%%%%%%%%%%%%%%%%%
\ba\label{eq:a 2G1G TC BS}
a_{\2G/\1G}^\tide\simeq\frac{Gm_{\1G}}{v_s^2}\frac{m_{\2G}}{m_s},
~~~a_{\2G/\1G}^\mathrm{BS}\simeq\frac{G m_{\1G}}{v_s^2}\bigg(\frac{m_{\2G}}{2m_s}\bigg)^{2/3}.
\ea
%%%%%%%%%%%%%%%%%%%%%%%%%%%%%%%%%%%%%%%%%%%%%%%%%%%%%%%%%%%%%%%%%%%%%%
As with 2G/2G binaries, these 2G/1G systems are generally soft and vulnerable to disruption by stellar scatterings,
strongly limiting their merger efficiency.

We compute the 1G/1G merger rates following the same order-of-magnitude estimates as for the 2G/2G case,
by substituting 2G BHs with their 1G counterparts, i.e.,
%%%%%%%%%%%%%%%%%%%%%%%%%%%%%%%%%%%%%%%%%%%%%%%%%%%%%%%%%%%%%%%%%%%%%%
\ba\label{eq:a 1G1G TC BS}
a_{\1G/\1G}^\tide\simeq\frac{Gm_{\1G}^2}{m_s v_s^2},
~~~a_{\1G/\1G}^\mathrm{BS}\simeq\frac{G m_{\1G}}{v_s^2}\bigg(\frac{m_{\1G}}{2m_s}\bigg)^{2/3}.
\ea
%%%%%%%%%%%%%%%%%%%%%%%%%%%%%%%%%%%%%%%%%%%%%%%%%%%%%%%%%%%%%%%%%%%%%%

The middle and bottom panels of Fig. \ref{fig:Merger rate} show that for $M_\bullet=4\times10^6M_\odot$,
the binary–single channel produces few 2G/1G mergers.
This is because, even though $a_{\2G/\1G}^\mathrm{BS}$ is comparable to $a_{\2G/\2G}^\mathrm{BS}$,
the lower total mass of the 2G/1G binary prolongs the GW merger timescale, thereby yielding $T_\mathrm{m}/T_\mathrm{evap}\gg1$.
In contrast, the tidal capture channel dominates 2G/1G production due to the high 1G BH density,
though most of these binaries also satisfy $T_\mathrm{m}/T_\mathrm{evap}\gg1$,
with only $f_3\lesssim15\%$ entering the ZLK window where $T_\mathrm{m}\lesssim T_\mathrm{evap}$.
For 1G/1G mergers, neither tidal capture nor binary–single channels are effective: tidal capture binaries are too compact
($a_{\1G/S}^\tide\ll a_{\2G/S}^\tide$)
and merge before further encounters,
while systems formed in the binary–single channel experience weak ZLK excitation ($a_{\1G/\1G}^\mathrm{BS}\ll a_{\2G/\2G}^\mathrm{BS}$).
Only GW capture contributes significantly to 1G/1G mergers
\footnote{As shown in Fig. \ref{fig:Merger rate} (bottom panel),
the 1G/1G merger rate from our dynamical-secular channels is negligibly low (orders of magnitude below the total LVK rate).
Therefore, the 2G BHs in our model cannot be produced by these particular channels.
Instead, they are assumed to be supplied by other well-established mechanisms
(e.g., isolated binary evolution, SMBH-induced ZLK mergers, or dynamical scatterings; and references therein).
This external supply justifies our steady-state approximation for $N_{\rm 2G}$ in the rate calculation.
As a consistency check, the 1G/1G merger rate implied by maintaining $N_{\rm 2G} = {\rm const}$ can be estimated as
$\Gamma \simeq N_{\rm 2G}/T_A$ (see Appendix \ref{App:Appendix B}).
For our fiducial value ${\cal F}_{\rm 2G}=0.05\%$, rescaling to the number of Milky Way-like galaxies in the local universe
yields a rate of a few $\mathrm{Gpc}^{-3}\,\mathrm{yr}^{-1}$.
This value represents the total 1G/1G merger rate within NSCs arising from various formation channels (not just those considered in this work),
and falls below the LVK-inferred total BHB merger rate of $27.5$-$49.4 \, \mathrm{Gpc}^{-3}\,\mathrm{yr}^{-1}$ \citep[e.g.,][]{LIGO-2026}.
Thus, the steady-state 2G BH abundance assumed in our model does not overproduce 1G/1G mergers relative to current observational constraints.
}.
As $M_\bullet$ increases, the influence radius $R_\mathrm{inf}$ expands \citep[e.g.,][]{Tremaine 2002},
shifting the active merger region outward and reducing merger rates.
For $M_\bullet\gtrsim5\times10^7M_\odot$, only the GW capture channel remains viable.
This is because the high velocity dispersion prolongs the encounter times, allowing binaries to merge before further interactions,
while the larger galactic distances suppress the ZLK effect.

\section{Limit on Hierarchical mergers}
\label{sec 5}

In the hierarchical merger scenario, 1G BHs normally form with low spins ($\chi\lesssim0.3$) from stellar collapse \citep[e.g.,][]{Fuller 2019}.
A 1G/1G merger produces a 2G BH; for comparable mass progenitors, the merger remnant has a spin $\simeq0.7$
and the recoil velocity typically below 300km/s, allowing its retention in NSCs.
However, mergers involving at least one high spin BH (e.g., 2G/1G or 2G/2G) yield 3G BHs with larger recoils.

\begin{figure}
\centering
\begin{tabular}{cccc}
\includegraphics[width=7.5cm]{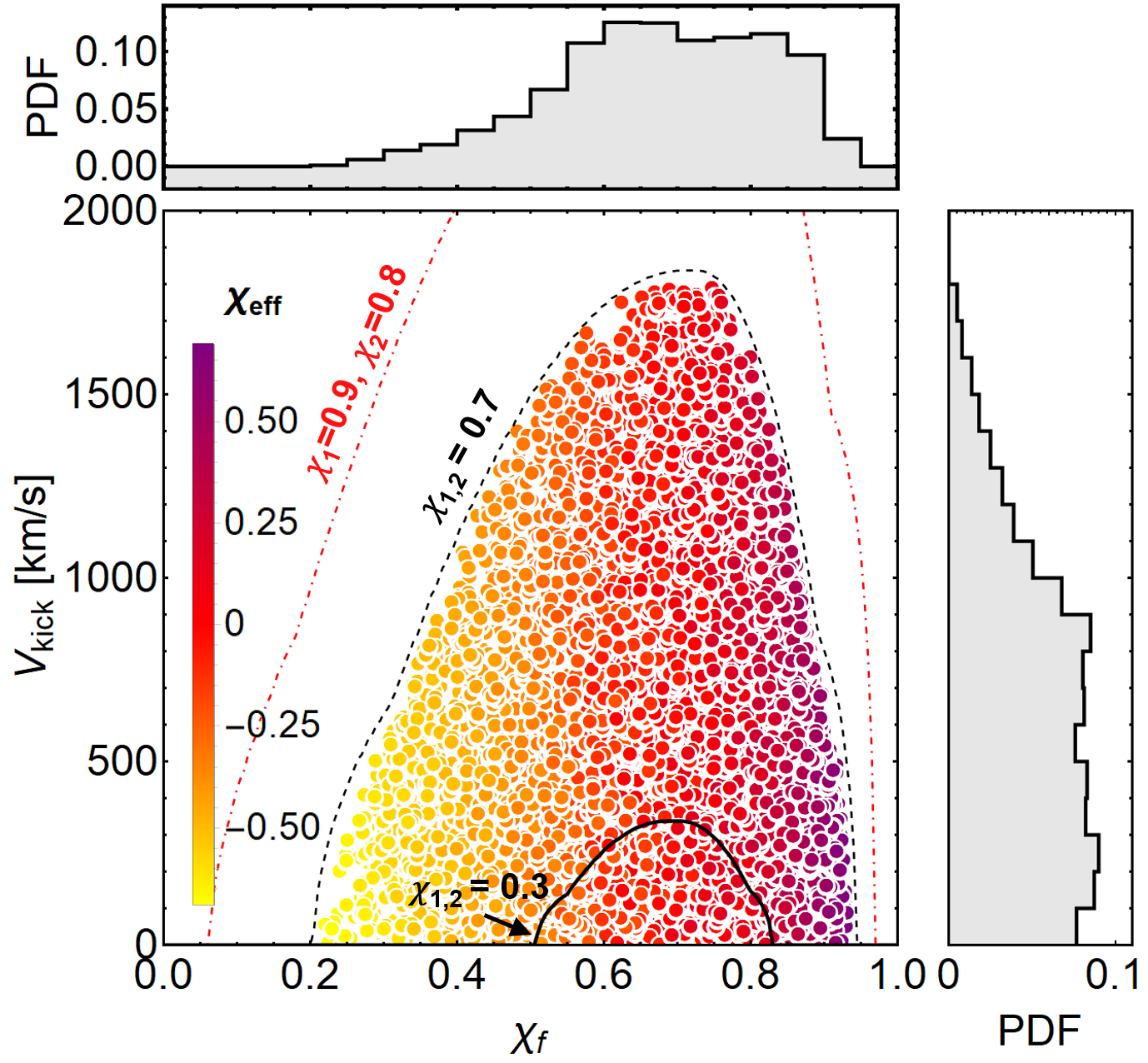}
\end{tabular}
\caption{Distribution of merger kick velocity versus final spin magnitude for equal-mass spinning BH mergers,
assuming isotropic spin orientations (see the analytical fits in \cite{Lousto 2010}).
Color indicates the value of effective spin parameter $\chi_\mathrm{eff}$,
defined as $\chi_{\rm eff}=(m_1 \chi_1 \theta_1+m_2 \chi_2 \theta_2)/(m_1+m_2)$,
where $\theta_{1,2}$ are the spin-orbit misalignment angles.
The black dashed contour shows the boundary of distribution for the case of $\chi_{1,2}=0.7$.
For reference, the black solid contour corresponds to low spin case ($\chi_{1,2}=0.3$),
and the red dot-dashed contours represent high unequal spins ($\chi_1=0.9$, $\chi_2=0.8$).
}
\label{fig:Merger kick}
\end{figure}

Fig. \ref{fig:Merger kick} shows the distribution of merger kick versus final spin.
As noted earlier, 2G/2G mergers in Milky Way-like galaxies primarily occur in the regions beyond 0.1 pc from the SMBH,
where the local escape velocity $v_s\lesssim300$ km/s.
For equal mass, equal spin binaries ($\chi_{1,2}=0.7$),
about $75\%$ of mergers have kicks above 300 km/s,
and over $58\%$ exceed 500 km/s.
Most 3G BH remnants are therefore ejected from their host clusters,
and their return through mass segregation is difficult, strongly suppressing further hierarchical growth.
This prediction offers a possible difference from environments (e.g., AGN disks) where higher-generation BHs may be retained.

\section{Conclusion and discussion}
\label{sec 6}

We have systematically investigated dynamical channels for 2G/2G mergers in nuclear star clusters,
considering four distinct populations with fixed masses for simplicity:
2G BHs (massive BHs in the high-mass gap), 1G BHs (low-mass BHs), single stars, and binary stars.
Our analysis suggests that 2G/2G mergers may predominantly form through sequential
binary–single interactions—beginning with 2G–star/star binary scattering, followed by 2G/star–star and 2G/star–2G encounters—that form 2G/2G binaries.
These binaries can then merge rapidly either through GW emission alone or via the SMBH-induced ZLK mechanism.
The resulting merger rate depends sensitively on the assumed 2G BH number density ${\cal F}_{\2G}$,
and can be consistent with that inferred for GW231123 within the uncertainties of our model.
This channel favors Milky Way–like galactic nuclei over more massive galaxies.

Nuclear star clusters are also efficient in producing 2G/1G binary mergers.
Their relatively weak GW \mbox{signals} or large distances may hinder detection,
but their long inspirals could make them promising low-frequency sources for future space-based detectors
like LISA, Tianqin, and Taiji \citep[e.g.,][]{LISA,TianQin,TaiJi}.
As a distinctive observational signature of our channel,
we further predict a significant population of 2G/star binaries,
whose mergers could appear as micro-tidal disruption events (micro-TDEs).
Detecting these systems would provide strong evidence for the proposed 2G/2G formation pathways.

Our results indicate that the 2G BHs in our channel originate primarily from isolated binary 1G/1G mergers,
where the 1G BHs themselves come from the direct collapse of massive stars, rather than from multiple generations of repeated mergers.
This is partly because low-mass 1G BHs merge inefficiently through typical dynamical processes,
and partly because hierarchical mergers produce high-spin BHs with large kicks, ejecting them and suppressing further growth.

This theoretical framework, while adopting a number of simplified assumptions
and intended as a proof of concept,
can be extended to systems with continuous mass spectra.
It may also provide physical insight into the origin of recent GW events such as GW241011 and GW241110 (which exhibit mass ratios $\sim2:1$ and large primary spins, consistent with 2G/1G mergers) \citep[e.g.,][]{GW241011}.

\section{Acknowledgments}

Bin Liu thanks Wenbin Lu and Johan Samsing for useful discussions.
Bin Liu acknowledges support from the National Natural Science Foundation of China (Grant No. 12433008)
and National Key Research and Development Program of China (No. 2023YFB3002502).

\appendix
\section{A:N-body SCATTERING EXPERIMENTS FOR BINARY-SINGLE ENCOUNTERS}
\label{App:Appendix A}

In channel 3, the formation of 2G/2G BHBs is primarily driven by two dynamical processes: 2G-S/S and 2G/S-2G encounters,
both of which involve an exchange interaction.
In the main text, we provided order-of-magnitude estimates for the semi-major axes and eccentricities of the resulting 2G/S and 2G/2G binaries.
Here, we perform a series of N-body scattering simulations to obtain the distributions of these orbital parameters.
By fitting the peak values of the distributions, we calibrate the prefactors in the analytical scaling relations.
We also quantify the probabilities of various encounter outcomes, including the formation fraction of the bound 2G/2G binaries.
Other three-body encounters that do not directly lead to 2G/2G formation, such as 2G/S-S or 2G/2G-S interactions, are omitted.
A detailed N-body investigation of the full set of processes is beyond the scope of this proof-of-concept work and is left for future study.

The system is configured as follows:
we consider a binary with component masses $m_1$ and $m_2$, semi-major axis $a_{\rm in}$, and eccentricity $e_{\rm in}$.
The Keplerian orbital velocity of the binary is $v_{\rm in} = \sqrt{G(m_1+m_2)/a_{\rm in}}$.
An external third body of mass $m_3$ approaches from infinity with velocity $v_{\rm inf}$
on a nearly parabolic trajectory and interacts with the binary at pericenter distance $r_{\rm int}^{\rm BS}$.
The outcome of the scattering is thus primarily determined by three dimensionless parameters:
$v_{\rm inf} / v_{\rm in}$, $e_{\rm in}$, and $\overline{r_\Int^{\rm BS}}$ (i.e., the normalized pericenter distance; see below).

For the 2G-S/S encounter, the process is analogous to the Hill mechanism.
During the interaction, the binary star (S/S) can be disrupted, typically producing a bound 2G/S binary while ejecting the other star.
The tidal disruption radius is given by $r_{\rm tide} \simeq a_{s/s} [m_{\rm 2G}/(2m_s)]^{1/3}$.
We assume a dimensionless interaction distance of
%%%%%%%%%%%%%%%%%%%%%%%%%%%%%%%%%%%%%%%%%%%%%%%%%%%%%%%%%%%%%%%%%%%%%%
\be
\overline{r_{\rm int}^{\rm BS}} \equiv \frac{r_{\rm int}^{\rm BS}}{r_{\rm tide}}\bigg|_{2G-S/S} = 1.
\ee
%%%%%%%%%%%%%%%%%%%%%%%%%%%%%%%%%%%%%%%%%%%%%%%%%%%%%%%%%%%%%%%%%%%%%%
Based on the analytical calculations provided in the main text,
the semi-major axis of the newly formed 2G/S binary is given by
%%%%%%%%%%%%%%%%%%%%%%%%%%%%%%%%%%%%%%%%%%%%%%%%%%%%%%%%%%%%%%%%%%%%%%
\be\label{eq: a2GS fitting}
a_{\2G/S}^\mathrm{BS} \simeq a_{s/s} \left( \frac{m_{\rm 2G}}{2m_s} \right)^{2/3} = \frac{G m_s}{v_s^2} \left( \frac{m_{\rm 2G}}{2m_s} \right)^{2/3},
\ee
%%%%%%%%%%%%%%%%%%%%%%%%%%%%%%%%%%%%%%%%%%%%%%%%%%%%%%%%%%%%%%%%%%%%%%
and the orbital eccentricity is
%%%%%%%%%%%%%%%%%%%%%%%%%%%%%%%%%%%%%%%%%%%%%%%%%%%%%%%%%%%%%%%%%%%%%%
\be\label{eq: e2GS fitting}
e_{\2G/S}^\mathrm{BS}\simeq 1 - \left( \frac{2m_s}{m_{\rm 2G}} \right)^{1/3}.
\ee
%%%%%%%%%%%%%%%%%%%%%%%%%%%%%%%%%%%%%%%%%%%%%%%%%%%%%%%%%%%%%%%%%%%%%%
In our N-body integrations, we set $m_1 = m_2 = 1 M_\odot$, $a_{\rm in}=a_{s/s} = 0.03$ AU,
and $e_{\rm in}=e_{s/s} = 0$. The third body is a 2G BH with mass $100 M_\odot$.
For simplicity, the 2G BH has an initial velocity of $v_{\rm inf} = 0$ at infinity.
We perform $5.12 \times 10^5$ scattering simulations, evolving the system through a single pericenter passage.
The outcomes reveal that approximately $13\%$ of the encounters result in flybys,
$\sim47\%$ lead to the temporary capture of the 2G BH into a transient, bound triple system,
and $40\%$ undergo an exchange interaction that successfully forms bound 2G/S binaries.

\begin{figure}
\centering
\begin{tabular}{cccc}
\includegraphics[width=8cm]{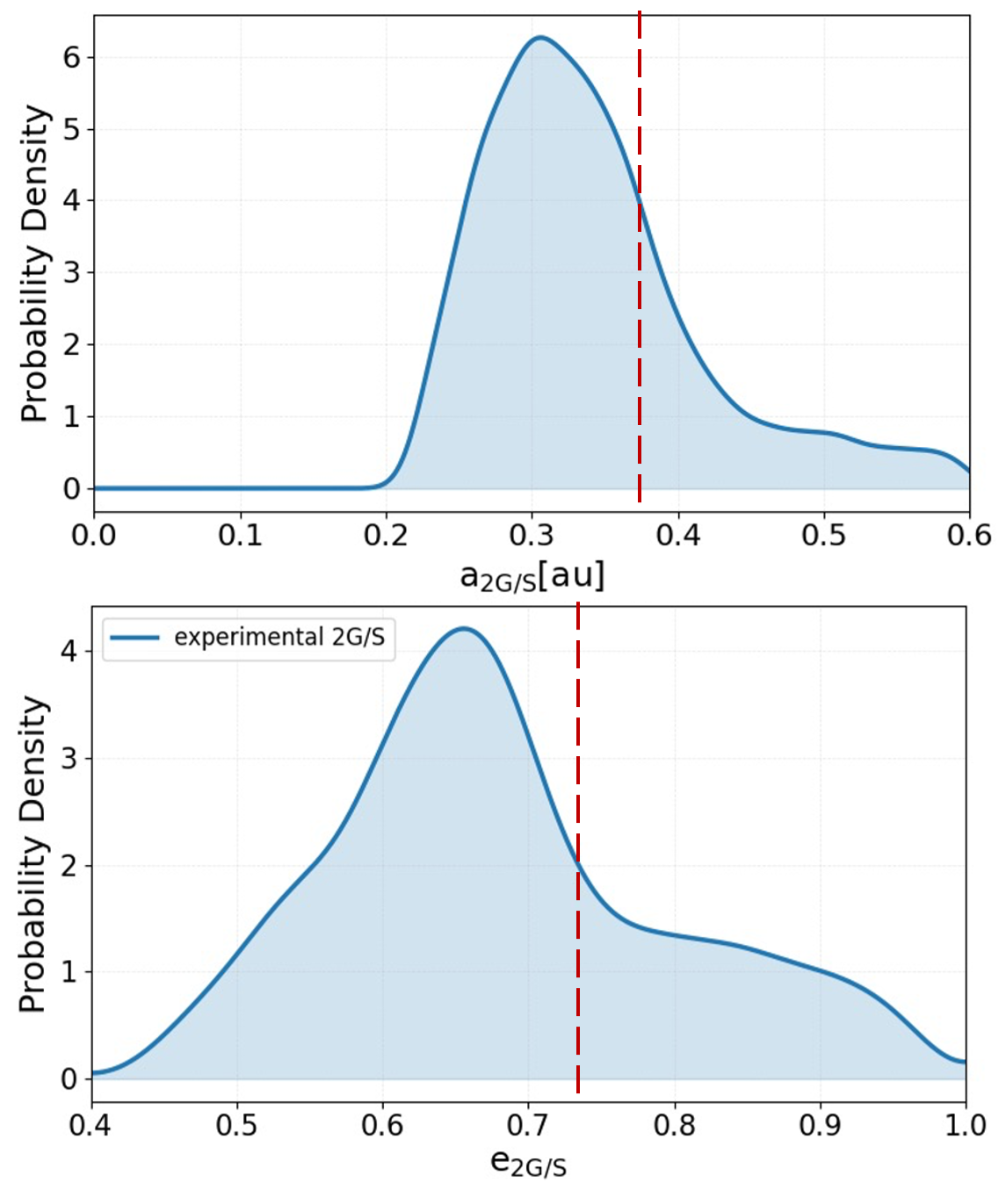}
\end{tabular}
\caption{Distributions of the semi-major axis $a_{\2G/S}^\mathrm{BS}$ and eccentricity $e_{\2G/S}^\mathrm{BS}$ of the newly formed 2G/S binaries
obtained from the N-body scattering simulations.
The vertical red dashed lines indicate the theoretical predictions from the analytical estimates
(Equations \ref{eq: a2GS fitting}-\ref{eq: e2GS fitting}).
}
\label{fig:scatter_results 2G-S/S}
\end{figure}

Figure~\ref{fig:scatter_results 2G-S/S} shows the distributions of
$a_{\2G/S}^\mathrm{BS}$ and $e_{\2G/S}^\mathrm{BS}$ obtained from our numerical simulations.
We find that both parameters are tightly constrained, with their peaks matching the theoretical predictions to within $30\%$.\\
\\
The 2G/S binary subsequently undergoes hardening through interactions with surrounding stars, with its semi-major axis decreasing to
\begin{equation}
a_{\2G/S}^{'\mathrm{BS}} = \zeta a_{\2G/S}^\mathrm{BS},
\end{equation}
where $\zeta$ is the hardening factor.
This hardening process continues until the 2G/S binary encounters another 2G BH.
The dimensionless pericenter distance for this subsequent encounter is defined as
\begin{equation}
\overline{r_{\rm int}^{\rm BS}} \equiv \frac{r_{\rm int}^{\rm BS}}{a_{\2G/S}^{'\mathrm{BS}}}\bigg|_{2G/S-S} = 1.
\end{equation}
For such a 2G/S-2G encounter, a 2G/2G BHB may form, with its semi-major axis given by
\ba\label{eq: a2G2G fitting}
&&a_{\2G/\2G}^\mathrm{BS} \simeq \eta a_{\2G/S}^{'\mathrm{BS}} \left(\frac{m_{\rm 2G}}{m_s}\right)\\
&&~~~~~~~~~= \zeta\eta \left[ a_{\2G/S}^\mathrm{BS} \left(\frac{m_{\rm 2G}}{m_s}\right) \right]\nonumber\\
&&~~~~~~~~~= \zeta\eta \left[ \frac{G m_{\rm 2G}}{v_s^2} \left(\frac{m_{\rm 2G}}{2m_s}\right)^{2/3} \right]\nonumber,
\ea
and its eccentricity given by
\begin{equation}
e_{\2G/\2G}^\mathrm{BS} \simeq 1 - \left(\frac{m_s}{m_{\rm 2G}}\right).
\end{equation}

In our N-body simulations, the target 2G/S binary consists of two components with masses
$m_1 = m_{\rm 2G} = 100 M_\odot$ and $m_2 = m_s = 1 M_\odot$.
The initial values of $a_{\2G/S}^\mathrm{BS}$ and $e_{\2G/S}^\mathrm{BS}$ are extracted
from the peak values of the distributions resulting from the prior 2G-S/S scatterings.
The incoming 2G BH approaches from infinity with $v_{\rm inf} = 0$ on a parabolic trajectory.
For the interaction distance $r_{\rm int}^{\rm BS}=\zeta a_{\2G/S}^\mathrm{BS}$,
we consider different hardening factors: $\zeta = 0.3$, $0.5$, and $0.8$.
For each case, we perform $5.12 \times 10^5$ scattering experiments and record the probabilities of the different outcomes
(shown in Table~\ref{TB: fractions}).

\begin{table}[htbp]
\centering
\caption{Fitted values of $\eta$ in Equation (\ref{eq: a2G2G fitting})
and the outcome probabilities from the scattering simulations for each $\zeta$.}
\label{TB: fractions}
\begin{tabular}{ccccc}
\hline
$\zeta$ & $\eta$ & $f_{\rm 2G/S-2G}$ & $f_{\rm 2G/2G-S}$ & $f_{\rm 2G/2G/S}$ \\
\hline
$0.5$ & $0.33$ & $26.9\%$ & $31.5\%$ & $41.6\%$ \\
$0.3$ & $0.33$ & $26.5\%$ & $30.7\%$ & $42.8\%$ \\
$0.8$ & $0.31$ & $25.9\%$ & $30.8\%$ & $43.3\%$ \\
\hline
\end{tabular}
\end{table}

\begin{figure}
\centering
\begin{tabular}{cccc}
\includegraphics[width=8cm]{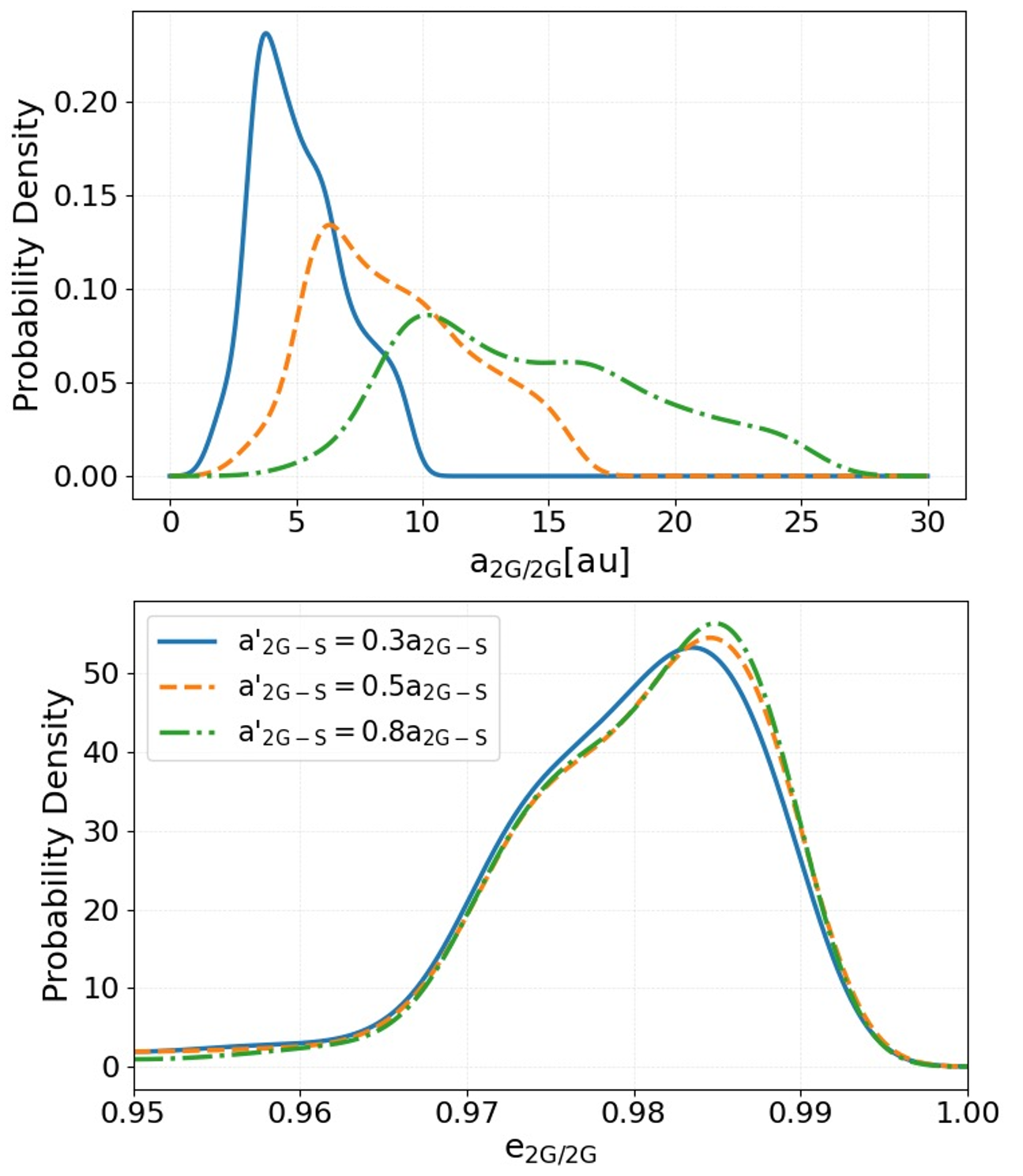}
\end{tabular}
\caption{Outcomes of the N-body scattering simulations for (2G/S)--2G encounters.
The top and bottom panels display the post-encounter distributions of the semi-major axis $a_{\2G/\2G}^\mathrm{BS}$
and the eccentricity $e_{\2G/\2G}^\mathrm{BS}$ for three different hardening factors (as labeled).
}
\label{fig:scatter_results 2G/S-S}
\end{figure}

Figure~\ref{fig:scatter_results 2G/S-S} shows the distributions of $a_{\2G/\2G}^\mathrm{BS}$
and $e_{\2G/\2G}^\mathrm{BS}$ obtained from our N-body integrations.
The semi-major axis $a_{\2G/\2G}^\mathrm{BS}$ exhibits a broad distribution, with the minimum and maximum values differing by a factor of $\sim5$.
From the peak of the distribution, we obtain $\eta \simeq 1/3$ , which holds for all considered values of $\zeta$ considered
\footnote{
For the merger rate calculation of 2G/2G binaries, we also examine the extreme values of $a_{\2G/\2G}^\mathrm{BS}$ for the case where $\zeta = 1/2$.
In this case, the minimum and maximum values thereby correspond to $\eta\simeq1/6$ and $\eta\simeq3/4$, respectively.
}.
The eccentricity $e_{\2G/\2G}^\mathrm{BS}$ is highly concentrated near 0.99, in agreement with the analytical estimate.

Note that 2G/S-2G encounters can lead to four distinct outcomes:
(i) flybys (2G/S-2G);
(ii) the formation of a 2G/2G binary accompanied by the ejection of the star (2G/2G-S);
(iii) the temporary capture of the incoming 2G BH to form a bound triple (2G/2G/S); and
(iv) the disruption of the entire system (2G-2G-S).
Table~\ref{TB: fractions} lists the probabilities of the first three outcomes,
whereas the fourth case has a zero probability ($f_{\rm 2G-2G-S}=0$) due to the vanishing velocity at infinity ($v_{\rm inf} = 0$).
We see that the probability of forming a 2G/2G binary, denoted as $f_{\rm 2G/2G-S} = f_2$ in the merger rate calculation of the main text,
is found to be approximately $30\%$ across our simulations.

\begin{figure}
\centering
\begin{tabular}{cccc}
\includegraphics[width=8cm]{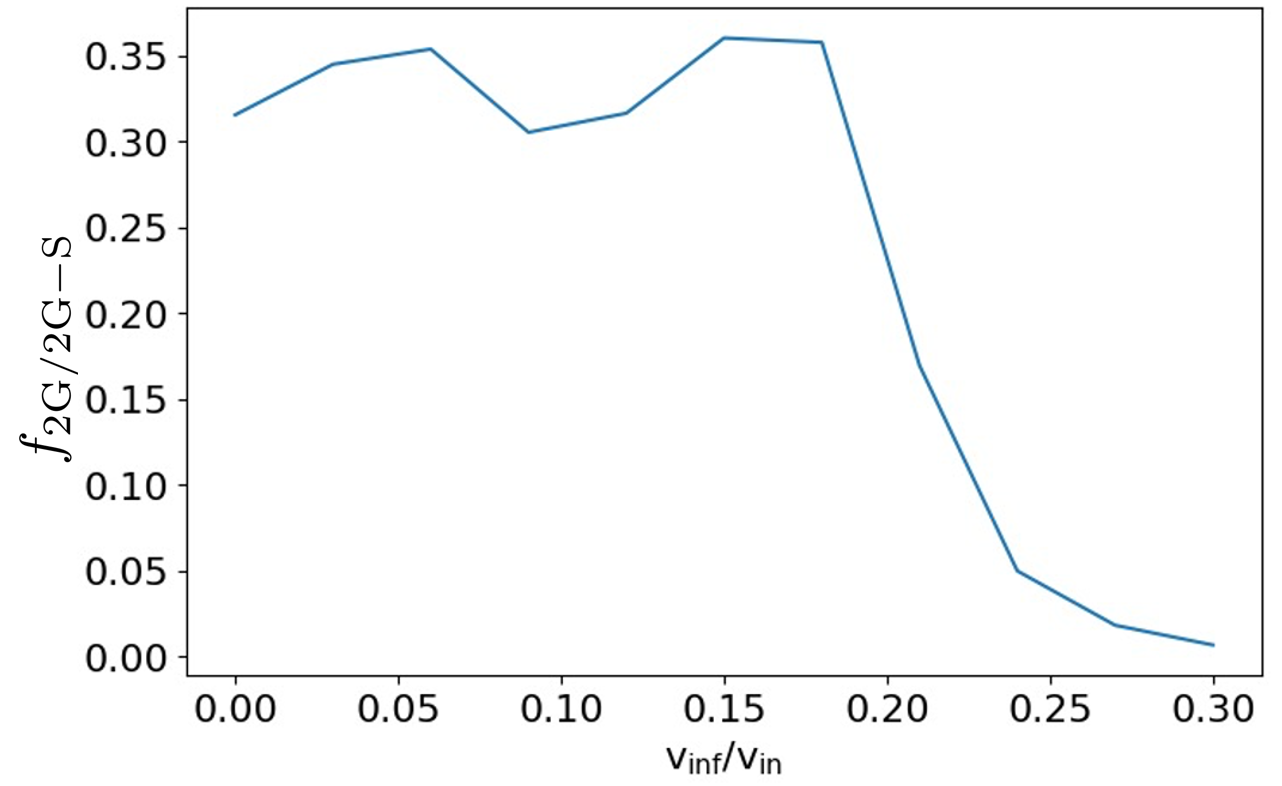}
\end{tabular}
\caption{The 2G/2G binary formation probability $f_{\rm 2G/2G-S}$ as a function of the normalized incoming velocity $v_{\rm inf}/v_{\rm in}$.
}
\label{fig:f2_sensitivity}
\end{figure}

We also explore the sensitivity of $f_{\rm 2G/2G-S}$ to the incoming velocity $v_{\rm inf}$ in our scattering simulations,
which can vary the total energy of the system.
The results are shown in Figure~\ref{fig:f2_sensitivity}.
For $v_{\rm inf}/v_{\rm in}\lesssim0.18$, $f_{\rm 2G/2G-S}$ remains approximately $30\%$ and
then rapidly drops to zero as $v_{\rm inf}$ increases further.

\section{B: Merger rate}
\label{App:Appendix B}

This section presents the merger rate calculations used in the main text for a typical galaxy.
For the GW capture channel, 2G BHs evolve through three stages into 3G BHs (see Fig. 1 in the main text).
We track three ``intermediate populations" as:
(1) single 2G BHs, $N_{\2G}$; (2) 2G/2G binary, $N_{\2G/\2G}$; and (3) 3G BHs, $N_{\3G}$.
Their evolution is described by:
%%%%%%%%%%%%%%%%%%%%%%%%%%%%%%%%%%%%%%%%%%%%%%%%%%%%%%%%%%%%%%%%%%%%%%
\[
\begin{dcases}
\frac{dN_{\2G}}{dt}=-\frac{N_{\2G}}{T_{\2G-\2G}}+\bigg(\frac{dN_{\2G}}{dt}\bigg)_{source}\simeq0\\
\frac{dN_{\2G/\2G}}{dt}=\frac{N_{\2G}}{T_{\2G-\2G}}-\frac{N_{\2G/\2G}}{T_\mathrm{m}}\\
\frac{dN_{\3G}}{dt}=\frac{N_{\2G/\2G}}{T_\mathrm{m}},
\end{dcases}
\]
%%%%%%%%%%%%%%%%%%%%%%%%%%%%%%%%%%%%%%%%%%%%%%%%%%%%%%%%%%%%%%%%%%%%%%
where the initial value $N_{\2G}(0)=N_{\2G 0}\neq0$, $N_{\2G/\2G}(0)=N_{\3G}(0)=0$.
The solution is:
%%%%%%%%%%%%%%%%%%%%%%%%%%%%%%%%%%%%%%%%%%%%%%%%%%%%%%%%%%%%%%%%%%%%%%
\[
\begin{dcases}
N_{\2G}(t)=N_{\2G 0}\\
N_{\2G/\2G}(t)=\frac{N_{\2G 0}T_\mathrm{m}}{T_{\2G-\2G}}\bigg(1-e^{-t/T_\mathrm{m}}\bigg).
\end{dcases}
\]
%%%%%%%%%%%%%%%%%%%%%%%%%%%%%%%%%%%%%%%%%%%%%%%%%%%%%%%%%%%%%%%%%%%%%%
The resulting merger rate of 2G/2G BHB at distance $R$ is given by:
%%%%%%%%%%%%%%%%%%%%%%%%%%%%%%%%%%%%%%%%%%%%%%%%%%%%%%%%%%%%%%%%%%%%%%
\be\label{eq: merger rate GW}
\Gamma=\frac{dN_{\3G}}{dt}
=\frac{dN_{\2G 0}}{dlnR}\frac{1}{T_{\2G-\2G}}\bigg(1-e^{-\frac{T_\mathrm{H}}{T_\mathrm{m}}}\bigg).
\ee
%%%%%%%%%%%%%%%%%%%%%%%%%%%%%%%%%%%%%%%%%%%%%%%%%%%%%%%%%%%%%%%%%%%%%%
Here, we set $t=T_\mathrm{H}=10^{10}$yrs.

For the tidal capture and binary–single interaction channels, as mentioned in the main text,
an additional stage of 2G/S binary formation is included, leading to the following extended set of equations:
%%%%%%%%%%%%%%%%%%%%%%%%%%%%%%%%%%%%%%%%%%%%%%%%%%%%%%%%%%%%%%%%%%%%%%
\[
\begin{dcases}
\frac{dN_{2G}}{dt}=-\frac{N_{2G}}{T_A}+\big(\frac{dN_{2G}}{dt}\big)_{\mathrm{source}}\simeq0\\
\frac{dN_{2G/S}}{dt}=f_1\frac{N_{2G}}{T_A}-\frac{N_{2G/S}}{T_B}\\
\frac{dN_{2G/2G}}{dt}=f_2\frac{N_{2G/S}}{T_B}-\frac{N_{2G/2G}}{T_{\mathrm{m}}}\\
\frac{dN_{3G}}{dt}=f_3\frac{N_{2G/2G}}{T_{\mathrm{m}}}.
\end{dcases}
\]
%%%%%%%%%%%%%%%%%%%%%%%%%%%%%%%%%%%%%%%%%%%%%%%%%%%%%%%%%%%%%%%%%%%%%%
where $T_A\equiv T_{\2G-S}$(tidal capture) or $T_{\2G-S/S}$ (binary single interaction),
and $T_B\equiv T_{\2G/S-\2G}$.
Since $N_{\2G}$ is a constant, we can find the analytical solution as follows:
%%%%%%%%%%%%%%%%%%%%%%%%%%%%%%%%%%%%%%%%%%%%%%%%%%%%%%%%%%%%%%%%%%%%%%
\[
\begin{dcases}
N_{\2G}(t)=N_{\2G 0}\\
N_{\2G/S}(t)=f_1\frac{N_{\2G 0}T_B}{T_A}\bigg(1-e^{-\frac{t}{T_B}}\bigg)\\
N_{\2G/\2G}(t)=f_1 f_2\frac{N_{\2G 0} T_{\mathrm{m}}}{T_A}
\bigg(1+\frac{T_{\mathrm{m}} e^{-\frac{t}{T_{\mathrm{m}}}}-
T_B  e^{-\frac{t}{T_B}}}{T_B - T_{\mathrm{m}}} \bigg)
\end{dcases}
\]
%%%%%%%%%%%%%%%%%%%%%%%%%%%%%%%%%%%%%%%%%%%%%%%%%%%%%%%%%%%%%%%%%%%%%%

\begin{figure}
\centering
\begin{tabular}{cccc}
\includegraphics[width=8cm]{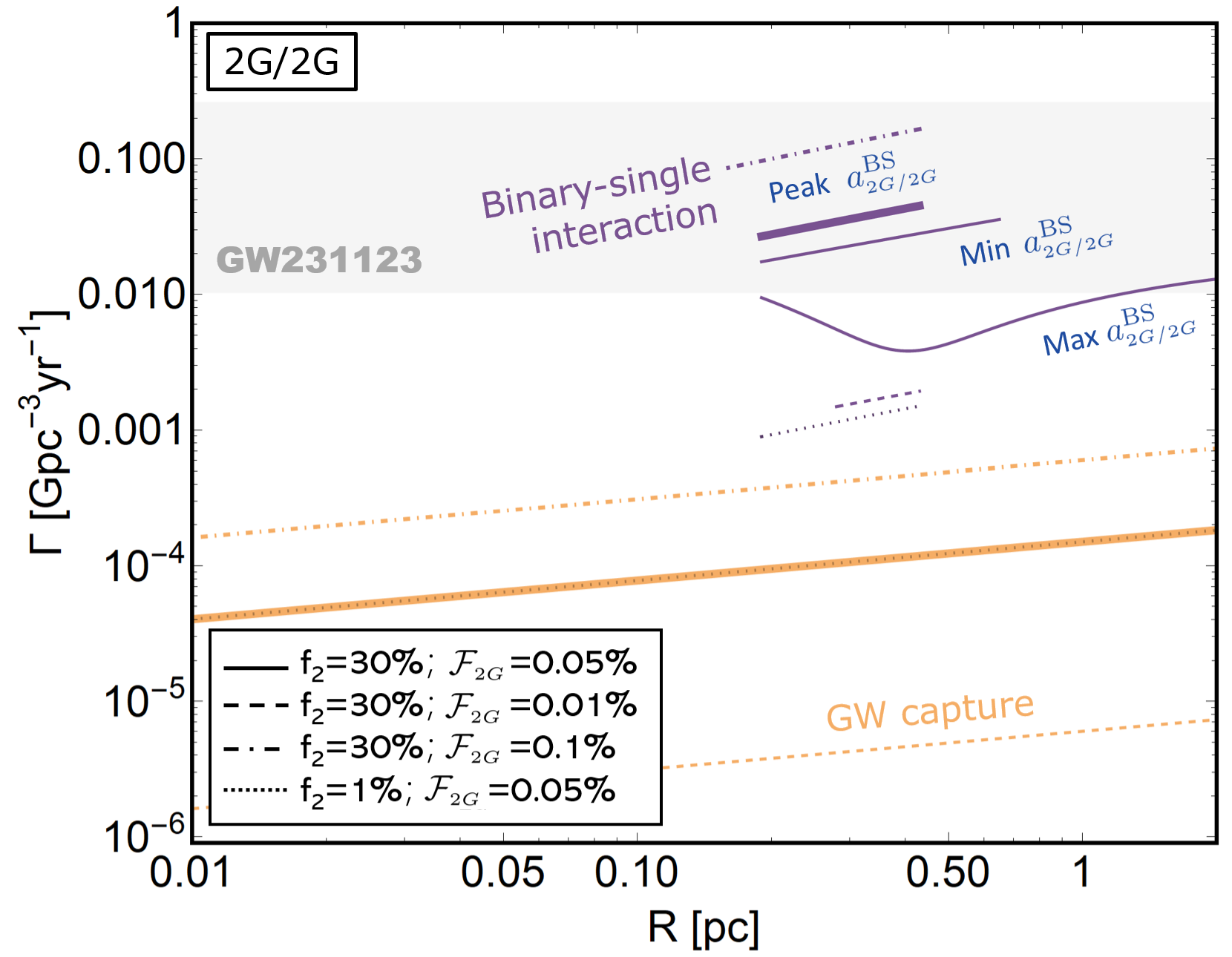}
\end{tabular}
\caption{Merger rate of 2G/2G binaries versus galactocentric distance $R$.
Since Eqs. (\ref{eq: merger rate GW}) and (\ref{eq: merger rate BS}) give the merger rate per galaxy,
we rescale the rate to Gpc$^{-3}$yr$^{-1}$ using the number density of Milky Way-like galaxies for comparison with the GW231123 detection rate
($0.08^{+0.19}_{-0.07}$Gpc$^{-3}$yr$^{-1}$; Gray shaded region) \citep[e.g.,][]{GW231123,SDSS 2003,Eagle 2014}.
Different values of $f_2$ and ${\cal F}_{\2G}$ are considered as labeled.
For the GW capture channel, $T_\mathrm{m}$ in Eq. (\ref{eq: merger rate GW}) is the merger time of an isolated binary \citep[e.g.,][]{Peters}.
For the binary–single channel, ZLK oscillations induced by the SMBH can accelerate 2G/2G mergers,
especially when $a_{\2G/\2G}^\mathrm{BS}$ reaches the upper tail of its distribution.
$T_\mathrm{m}$ in Eq. (\ref{eq: merger rate BS}) is set to $T_\mathrm{evap}$, and
$f_3$ denotes the numerically evaluated fraction of systems with $T_\mathrm{m}\lesssim T_\mathrm{evap}$ \citep[e.g.,][]{Liu-HierarchicalMerger}.
}
\label{fig:Different f}
\end{figure}

Then, for a given distance $R$, the merger rate is
%%%%%%%%%%%%%%%%%%%%%%%%%%%%%%%%%%%%%%%%%%%%%%%%%%%%%%%%%%%%%%%%%%%%%%
\ba\label{eq: merger rate BS}
\Gamma
=\frac{dN_{\2G 0}}{dlnR}\frac{f_1 f_2 f_3}{T_A}
\bigg(1+\frac{T_{\mathrm{m}} e^{-\frac{T_\mathrm{H}}{T_{\mathrm{m}}}}-
T_B  e^{-\frac{T_\mathrm{H}}{T_B}}}{T_B - T_{\mathrm{m}}} \bigg).
\ea
%%%%%%%%%%%%%%%%%%%%%%%%%%%%%%%%%%%%%%%%%%%%%%%%%%%%%%%%%%%%%%%%%%%%%%

The parameters $f_1$ and $f_2$ in Eq.~(\ref{eq: merger rate BS}) represent the probabilities of forming a 2G/S binary via a 2G–S/S encounter
and assembling a 2G/2G binary through a subsequent interaction with another 2G BH, respectively.
The values depend on the three-body mass and orbital properties during the encounter,
and should ideally be determined via N-body simulations.
The parameter ${\cal F}_{\2G}=n_s/n_{\2G}$ denotes the abundance ratio of 2G BHs to stars,
which can vary the encounter timescale.
For the fiducial example in the main text, we adopt fixed values of $f_1=40\%$, $f_2=30\%$ and ${\cal F}_{\2G}=0.05\%$.

As shown in Fig. \ref{fig:Different f}, we examine the dependence of $\Gamma$ on these parameters.
We see that the merger rate $\Gamma$ depends linearly on $f_2$ for the binary–single channel, while it is highly sensitive to ${\cal F}_{\2G}$.
Given the broad distribution of $a_{\2G/\2G}^\mathrm{BS}$ (see Appendix \ref{App:Appendix A}),
we consider both the minimum and maximum values. The lower bound behaves similarly to the peak case,
while for the upper bound, the ZLK mechanism plays an important role in producing 2G/2G binary mergers.
To compare with the observed rate of GW231123,
we rescale $\Gamma$ by considering the number of Milky Way-like galaxy within a $\lesssim$4Gpc$^3$ volume (i.e., $\sim10^7$)
\citep[e.g.,][]{SDSS 2003,Eagle 2014}.
We find that such channel is able to predict a merger rate in agreement with the detection rate of GW231123
when the 2G BH fraction ${\cal F}_{\2G}\gtrsim0.05\%$.

\section{C: Mergers from 2G/1G binaries}
\label{App:Appendix C}

In the following, the formation of 2G/1G binaries is characterized
via dynamical arguments relying solely on order-of-magnitude estimates, without being calibrated by N-body simulations.

\textit{Channel 1: Gravitational wave capture}.
This channel follows a similar dynamical process as the 2G/2G case.
Consider a gravitationally focused encounter between a 2G BH and a 1G BH,
with pericenter distance $r_\Int^\gw$.
The timescale for such encounters is given by
%%%%%%%%%%%%%%%%%%%%%%%%%%%%%%%%%%%%%%%%%%%%%%%%%%%%%%%%%%%%%%%%%%%%%%
\ba
T_{{\2G}-{\1G}}\simeq v_s\big[2\pi G (m_{\2G}+m_{\1G})n_{\1G}r_\Int^\gw\big]^{-1}.
\ea
%%%%%%%%%%%%%%%%%%%%%%%%%%%%%%%%%%%%%%%%%%%%%%%%%%%%%%%%%%%%%%%%%%%%%%
During the close passage, energy is radiated via GWs.
The amount of energy dissipated is approximately
$\Delta E_\gw\simeq(85\pi/12\sqrt{2})[G^{7/2}\mu_{\2G,\1G}^2(m_{\2G}+m_{\1G})^{5/2}]/[c^5(r_\Int^\gw)^{7/2}]$.
A bound 2G/1G binary forms if the total energy after the encounter becomes negative, i.e.,
$(1/2)\mu_{\2G,\1G} v_{\2G}^2-\Delta E_\gw (r_\Int)=E_b\equiv-Gm_{\2G}m_{\1G}/(2a_{{\2G}/{\1G}})\lesssim0$.
Moreover, for the binary to survive \mbox{subsequent} stellar interactions without being disrupted,
its binding energy must satisfy $E_b\lesssim-m_s v_s^2/2$.
This defines the characteristic semi-major axis:
%%%%%%%%%%%%%%%%%%%%%%%%%%%%%%%%%%%%%%%%%%%%%%%%%%%%%%%%%%%%%%%%%%%%%%
\ba
a_{{\2G}/{\1G}}^\gw\simeq\frac{Gm_{\2G}m_{\1G}}{m_s v_s^2}.
\ea
%%%%%%%%%%%%%%%%%%%%%%%%%%%%%%%%%%%%%%%%%%%%%%%%%%%%%%%%%%%%%%%%%%%%%%
The interaction radius $r_\Int^\gw$ is determined by setting the GW energy loss comparable to the binding energy requirement
($\Delta E_\gw\simeq m_s v_s^2$):
%%%%%%%%%%%%%%%%%%%%%%%%%%%%%%%%%%%%%%%%%%%%%%%%%%%%%%%%%%%%%%%%%%%%%%
\ba
r_\Int^\gw\equiv \bigg(\frac{85\pi}{12\sqrt{2}}\bigg)^{2/7}
\bigg[\frac{c^2}{v_s^2}\frac{m_{\1G}^2(m_{\1G}+m_{\2G})^{1/2}}{m_s m_{\2G}^{3/2}}\bigg]^{2/7}R_g,
\ea
%%%%%%%%%%%%%%%%%%%%%%%%%%%%%%%%%%%%%%%%%%%%%%%%%%%%%%%%%%%%%%%%%%%%%%
where $R_g=Gm_{\2G}/c^2$ is the gravitational radius of the 2G BH.
The characteristic eccentricity of the newly formed binary can be determined using the relation
$a_{{\2G}/{\1G}}^\gw(1-e_{{\2G}/{\1G}}^\gw)\simeq r_\Int^\gw$.

\textit{Channel 2: Tidal capture}.
A 2G BH can also form a binary with a 1G BH through an intermediate tidal capture step.
In this two-step process, a 2G BH first tidally captures a star to form a 2G/S binary.
The timescale for such a 2G-S encounter follows the same formulation as in the 2G/2G case.
The characteristic semi-major axis of the resulting 2G/S binary is
%%%%%%%%%%%%%%%%%%%%%%%%%%%%%%%%%%%%%%%%%%%%%%%%%%%%%%%%%%%%%%%%%%%%%%
\ba
a_{\2G/S}^{\tide}\simeq\frac{Gm_{\2G}}{v_s^2}.
\ea
%%%%%%%%%%%%%%%%%%%%%%%%%%%%%%%%%%%%%%%%%%%%%%%%%%%%%%%%%%%%%%%%%%%%%%
Subsequent gravitational scatterings with passing stars harden this binary, reducing its semi-major axis to $a_{\2G/S}^{' \tide}\simeq a_{\2G/S}^{\tide}/2$.
A 2G/1G binary may then form via a binary-single exchange interaction when
a 1G BH approaches the hardened 2G/star binary within a distance $\sim a_{\2G/S}^{' \tide}$. The timescale for this encounter is
%%%%%%%%%%%%%%%%%%%%%%%%%%%%%%%%%%%%%%%%%%%%%%%%%%%%%%%%%%%%%%%%%%%%%%
\ba
T_{\2G/S-\1G}\simeq v_s\big[2\pi G (m_{\2G}+m_{\1G})n_{\1G} a_{\2G/S}^{' \tide}\big]^{-1}.
\ea
%%%%%%%%%%%%%%%%%%%%%%%%%%%%%%%%%%%%%%%%%%%%%%%%%%%%%%%%%%%%%%%%%%%%%%
During this interaction, an energy exchange of order $\Delta E\simeq Gm_{\2G}m_s/a_{\2G/S}^{' \tide}$ occurs.
If the total energy becomes negative, a 2G/1G binary forms,
ejecting the lighter star. The binding energy of the newly formed binary satisfies $G(m_{\2G}+m_{\1G})/(2a_{\2G/\1G})\simeq\Delta E/2$,
giving a characteristic semi-major axis:
%%%%%%%%%%%%%%%%%%%%%%%%%%%%%%%%%%%%%%%%%%%%%%%%%%%%%%%%%%%%%%%%%%%%%%
\ba
a_{\2G/\1G}^\tide\simeq\frac{Gm_{\1G}}{v_s^2}\frac{m_{\2G}}{m_s}.
\ea
%%%%%%%%%%%%%%%%%%%%%%%%%%%%%%%%%%%%%%%%%%%%%%%%%%%%%%%%%%%%%%%%%%%%%%
The corresponding eccentricity follows from $a_{\2G/\2G}^\tide(1-e_{\2G/\2G}^\tide)\simeq a_{\2G/S}^{' \tide}$.
The resulting 2G/1G binary is relatively soft and could be disrupted by subsequent scatterings.
We therefore only consider binaries that merge within the evaporation timescale $T_\mathrm{evap}$.

\textit{Channel 3: Binary-single interaction}.
A third channel for 2G/1G mergers arises from binary-single interactions,
particularly between a 2G BH and a stellar binary.
This process is analogous to the 2G/2G case,
except that the encounter between a 2G/S binary and another 2G BH is replaced with interactions between a 2G/S binary and a 1G BH.
The characteristic semi-major axis of the 2G/star binary formed through this channel is
%%%%%%%%%%%%%%%%%%%%%%%%%%%%%%%%%%%%%%%%%%%%%%%%%%%%%%%%%%%%%%%%%%%%%%
\ba
a_{\2G/S}^\mathrm{BS}\simeq\frac{G m_s}{v_s^2}\bigg(\frac{m_{\2G}}{m_s}\bigg)^{2/3}.
\ea
%%%%%%%%%%%%%%%%%%%%%%%%%%%%%%%%%%%%%%%%%%%%%%%%%%%%%%%%%%%%%%%%%%%%%%
After hardening via scattering with field stars to $a_{\2G/S}^{' \mathrm{BS}}\simeq a_{\2G/S}^\mathrm{BS}/2$,
the binary may encounter a 1G BH. During this interaction, an energy exchange of order
$\Delta E\simeq Gm_{\2G}m_s/a_{\2G/S}^{' \mathrm{BS}}$ occurs, potentially forming a 2G/1G binary.
If $G(m_{\2G}+m_{\1G})/(2a_{\2G/\1G})\simeq\Delta E/2$,
the characteristic semi-major axis of the resulting 2G/1G binary is:
%%%%%%%%%%%%%%%%%%%%%%%%%%%%%%%%%%%%%%%%%%%%%%%%%%%%%%%%%%%%%%%%%%%%%%
\ba
a_{\2G/\1G}\simeq\frac{G m_{\1G}}{v_s^2}\bigg(\frac{m_{\2G}}{m_s}\bigg)^{2/3}.
\ea
%%%%%%%%%%%%%%%%%%%%%%%%%%%%%%%%%%%%%%%%%%%%%%%%%%%%%%%%%%%%%%%%%%%%%%
As with the other channels, the resulting 2G/1G binary is subject to disruptive encounters,
and only those merging within $T_\mathrm{evap}$ contribute to the merger rate calculations.

As a consistency check for 2G/1G mergers,
the tidal capture channel, which dominates the 2G/1G rate in our model,
yields $\Gamma_{\rm 2G/1G} \sim 0.005$--$0.04 \, \mathrm{Gpc}^{-3}\,\mathrm{yr}^{-1}$ (based on the middle panel of Fig.~\ref{fig:Merger rate}).
This is below the LVK-inferred rate of $0.2$--$3.11 \, \mathrm{Gpc}^{-3}\,\mathrm{yr}^{-1}$ for the hierarchical merger subpopulation at $z=0.2$
\citep[e.g.,][]{LIGO-2026}, which includes both 2G/1G and 2G/2G mergers.
We therefore conclude that our model does not overproduce merger rates already constrained by LVK observations.

%%%%%%%%%%%%%%%%%%%%%%%%%%%%%%%%%%%%%%%%%%%%%%%%%%%%%%%%%%%%%%%%%%%%%%

\end{document}